\documentclass[twocolumn]{aastex631}

\newcommand{\vtwo}[1]{{#1}}

\defcitealias{Espinoza2015}{EJ15}
\defcitealias{Espinoza2016}{EJ16}
\defcitealias{Sing2026}{S26}

\usepackage{float}

\newcommand{\lt}{^^3c}
\newcommand{\gt}{^^3e}
\usepackage{showyourwork}

\shorttitle{More Science from Transit Data}
\shortauthors{Mercier, de Wit, \& Rackham}

\graphicspath{{./}{figures/}}

\begin{document}

\title{What's in Your Transit? A Path to More Science from Exoplanet Transit Data}

\author[0000-0002-0962-7585]{Samson J.\ Mercier}
\affiliation{Department of Earth, Atmospheric and Planetary Sciences, Massachusetts Institute of Technology, 77 Massachusetts Avenue, Cambridge, MA 02139, USA}

\author[0000-0003-2415-2191]{Julien de Wit}
\affiliation{Department of Earth, Atmospheric and Planetary Sciences, Massachusetts Institute of Technology, 77 Massachusetts Avenue, Cambridge, MA 02139, USA}

\author[0000-0002-3627-1676]{Benjamin V.\ Rackham}
\affiliation{Department of Earth, Atmospheric and Planetary Sciences, Massachusetts Institute of Technology, 77 Massachusetts Avenue, Cambridge, MA 02139, USA}
\affiliation{Kavli Institute for Astrophysics and Space Research, Massachusetts Institute of Technology, Cambridge, MA 02139, USA}

\correspondingauthor{Samson J.\ Mercier}
\email{merci228@mit.edu}

\begin{abstract}
\vtwo{Transit-depth measurements can remain less precise than expected from photon noise alone, because limb darkening and orbital parameters are inferred simultaneously with transit depth. We refer to this shortfall as the ``transit-depth precision problem'' and quantify it with an amplification factor, $A$, comparing the measured transit-depth uncertainty with baseline scatter. 
Using central-transit injection--retrieval tests informed by $\gt500{,}000$ MPS-ATLAS intensity profiles, we determine how limb-darkening prescriptions and priors affect both transit-depth precision and accuracy.
We find that limb-darkening coefficients are the dominant contributors to the excess uncertainty, followed by inclination and stellar density, through parameter correlations. In the configurations tested, the widely used quadratic law can produce low $A$ yet substantially biased transit depths and wavelength-dependent structure in otherwise flat transmission spectra. Third-order polynomial and fourth-order nonlinear laws provide a better precision-accuracy tradeoff, although all quantitative results remain conditional on the adopted stellar models and transit geometry.
Improved stellar models can increase observing efficiency by enabling tighter, accurate limb-darkening coefficient priors: in our fourth-order nonlinear tests, $10\%$--$20\%$ priors reduce transit-depth uncertainties by factors of $1.1$--$1.8$ relative to uniform priors, while stronger $1\%$ priors reduce them by $2.1$--$2.5$. Assuming uncertainty decreases as the inverse square root of the number of observed transits, these gains correspond to reaching the same target transit-depth precision with approximately $1$--$3$ and $4$--$6$ times fewer transits, respectively. We recommend testing multiple limb-darkening prescriptions and reporting wavelength-resolved sensitivity to limb-darkening assumptions.
}
\end{abstract}


\section{Introduction}
\label{sec:intro}
Since its launch, the James Webb Space Telescope (JWST) has ushered in a new era of exoplanet science. Thanks to the unprecedented precision of its data, we can probe in finer detail the structure and composition of exoplanetary atmospheres \citep[e.g.,][]{Xue2024, Welbanks2024} and investigate novel processes such as photochemistry \citep[e.g.,][]{Tsai2023, Dyrek2024}, atmosphere--interior interactions \citep[e.g.,][]{Tsai2021, Benneke2024}, and planetary oblateness \citep[\textit{i.e.}, internal processes; e.g.,][]{Berardo2022, Liu2025}.

Although instrumentation has improved drastically over the last decades, the models and frameworks used to interpret data have lagged behind. One example is our incomplete understanding of stars: stellar magnetic active regions (spots and faculae) consistently contaminate photometric and spectroscopic observations \citep[e.g.,][]{Dumusque2011a, Dumusque2011b, Meunier2017, Rackham2017, Rackham2018, Rackham2019, Rackham2023, Saba2025}. Our inability to model or mitigate these effects leads to biased measurements of the radial-velocity amplitude and transit depth from which the most basic exoplanet properties---mass and radius, respectively---are measured \citep[e.g.,][]{Rabus2009, Carter2011, Bonfils2013}. Similar biases can arise from other model limitations, such as with line shape parameterizations in opacity models \citep{Niraula2022, Niraula2023}. These model limitations propagate throughout our analyses and hamper our ability to draw meaningful conclusions on exoplanet characteristics. 

In this context, here we examine the information content of exoplanet transit data with the goal of identifying what targeted improvements to methods and models can enhance the scientific return of these observations. In \autoref{sec:LD Context} we summarize prior work on limb-darkening in the context of exoplanet transits and clarify the novelties that our study adds to this well-established body of work.
In \autoref{sec:TDPP}, we present the root cause for our investigation, the ``transit-depth precision problem'' (TDPP). In \autoref{sec:framework} we construct the injection--retrieval framework underlying our subsequent analyses, \vtwo{informing} our choice of limb-darkening \vtwo{parameterization} in the statistics of modern stellar models. In \autoref{sec:amp_factor_cause}, we use this framework to investigate the origins of the TDPP and quantify their respective contributions. In \autoref{sec:boosting_science} we discuss targeted mitigation strategies\vtwo{, quantify the substantial boost in scientific return the community could expect from their adoption, isolate the chromatic signature of inadequate limb-darkening prescription, and reconcile our findings with recent empirical results}. We conclude in \autoref{sec:conclusion}. The whole paper is reproducible and can be found by clicking the GitHub icon \href{https://github.com/SamMerc/Transit-Information-Content}{\GitHubIcon}.

\begin{figure*}[!ht]
    \script{Fig1_plot.py}
    \centering
    \includegraphics[trim = 0 05mm 0 2mm , width=0.95\textwidth]{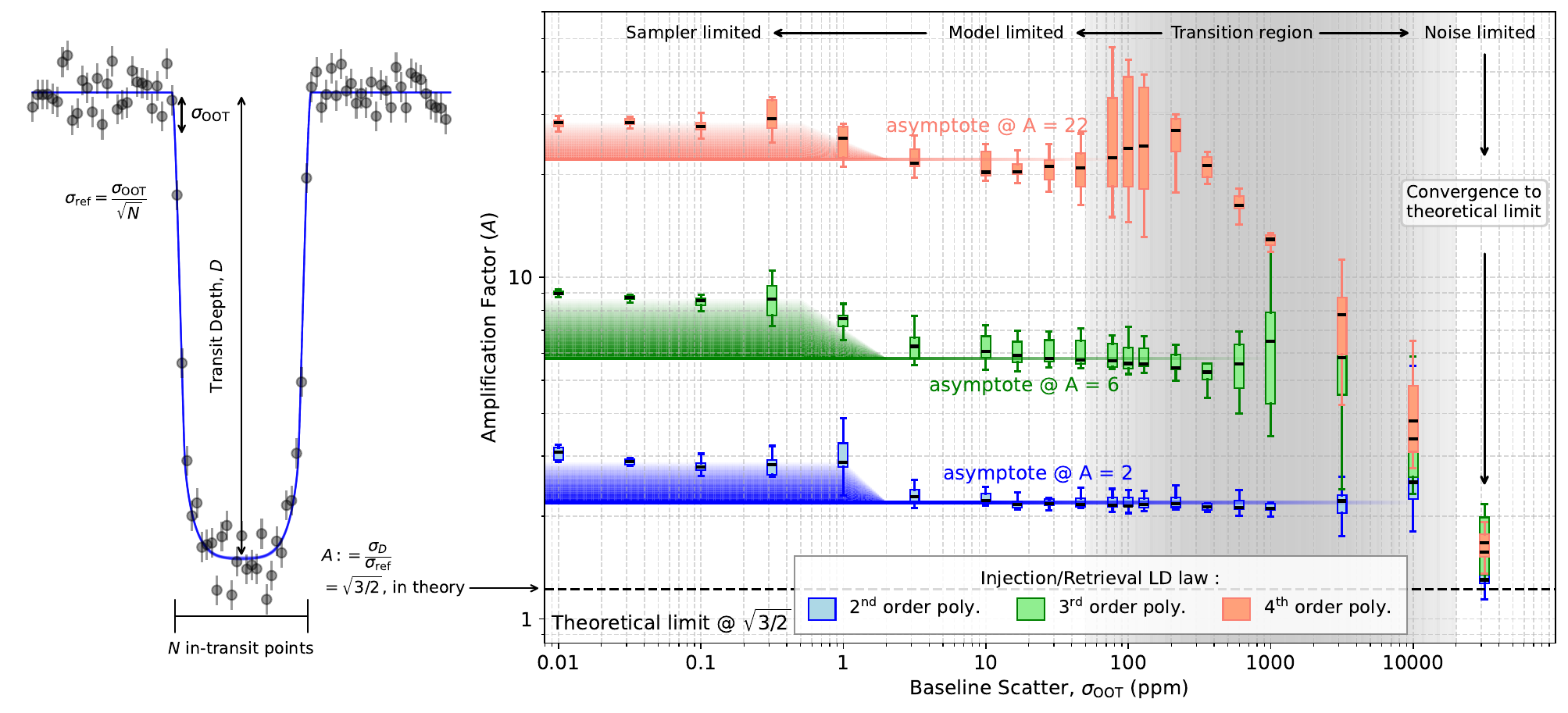}
    \caption{\textbf{Expectations vs.\ reality in constraining an exoplanet transit depth.}
    \textit{Left:} mock transit generated with the properties listed in \autoref{tab:tab1}. Black points show the light curve with 359 parts per million (ppm) of injected Gaussian white noise. Blue curves are model realizations drawn from the posterior distribution of $D$ obtained by fitting this dataset. Annotated arrows indicate the quantities relevant to the TDPP.
    \textit{Right:} illustration of the TDPP. Each colored box plot set corresponds to injection--retrieval tests done with a different limb-darkening prescription. For each set, the injection and retrieval models have the same functional form. Box plots summarize 10 Markov Chain Monte Carlo (MCMC) retrievals with identical scatter but different noise realizations. For each box, the black horizontal line marks the median, the lower and upper edges are the first and third quartiles, and whiskers are the minimum and maximum values (calculated as $1.5\times$ the first and third \vtwo{quartiles}). The four annotated regimes are discussed in \autoref{sec:TDPP}. Colored horizontal lines show the model-limited asymptote for each injection--retrieval family, fading toward the transition region. The injected out-of-transit scatter ($\sigma_{\rm OOT}$) is measured in 120 min bins.}
    \label{fig:fig1}
\end{figure*}

\section{Limb-darkening: prior work and open challenges}\label{sec:LD Context}
Limb-darkening, or center-to-limb intensity variation, occurs because radiation emerging near the stellar limb originates from higher, cooler atmospheric layers and is therefore weaker than radiation from the disk center. Since the advent of exoplanet transit observations \citep{Charbonneau2000,Henry2000}, limb-darkening has been recognized as a critical component of transit modeling, as it directly shapes the morphology of transit light curves.

Limb-darkening is typically described with parametric limb-darkening laws (LDLs), which express the stellar intensity as a function of the viewing angle $\mu$\footnote{$\mu = \cos(\theta)$ where $\theta$ is the angle between the observer’s line of sight and the stellar surface normal.} through a set of limb-darkening coefficients (LDCs). A wide variety of LDLs have been proposed, ranging from linear \citep{Milne1921} and quadratic forms \citep{Manduca1977} to higher-order polynomials \citep{Gimenez2006} and nonlinear prescriptions \citep{Claret2000, Sing2010}. In parallel, methodological advances have focused on reparameterizing LDLs to reduce parameter correlations \citep{Maxted2018,Maxted2023} and improve the sampling of LDC parameter spaces \citep{Kipping2013, Kipping2016}. 

This diversity of approaches reflects a fundamental tension between the intrinsic complexity of stellar intensity profiles, the increasing precision of transit observations, and the need for analytically and numerically tractable models. Stellar atmospheric models have long shown that realistic intensity profiles often require some degree of nonlinearity for accurate representation \citep[e.g.,][]{Claret1998, Neilson2013a, Neilson2013b}. Nevertheless, low-order LDLs remain widely used due to their simplicity and low dimensionality, which facilitate inference. 

An extensive body of work has demonstrated that such simplified LDLs lead to bias and increased variance in transit depth measurements \citep[e.g.,][]{Southworth2008, Carter2008, Csizmadia2013, Neilson2013a, Neilson2013b, Espinoza2015, Espinoza2016, Morello2017, Keers2024}. These effects arise from strong correlations between limb-darkening parameters and other transit parameters, which fundamentally limit the information that can be extracted from transit light curves. Such limitations become increasingly relevant for high-quality transit data, such as those acquired with JWST, where photon noise is low enough that modeling degeneracies dominate the error budget. Moreover, degraded transit depth precision and accuracy have been observed across a wide range of orbital geometries, wavelength regimes, and stellar types, underscoring that the choice of LDL has a ubiquitous impact on transit analyses. Many studies therefore recommend the use of higher-order laws, such as the three-parameter prescription of \cite{Sing2010}, to mitigate both bias and inflated variance. However, while higher-order LDLs reduce these effects, they do not eliminate them entirely, and the optimal choice of LDL remains strongly dependent on data quality \citep[see Figure 7 of][hereafter \citetalias{Espinoza2016}]{Espinoza2016}. \vtwo{Most recently, \citet[][hereafter \citetalias{Sing2026}]{Sing2026} pushed against this consensus. Comparing LDL choice on JWST transits of seven exoplanets orbiting FGK hosts, they found that laws more complex than the quadratic one are not warranted by the data, and that retrieved transit depths are negligibly biased by the use of the quadratic law, with respect to their Bayesian model average (BMA). Since these findings are in tension with the body of work above and our results, we will return to them in \autoref{sec:compare_empirics}}.

In summary, previous works have characterized limb-darkening effects in terms of bias, covariance structure, or goodness-of-fit, and explicitly quantified the information content of transit depth measurements relative to the photon-noise limit. However, no study has adopted a holistic Markov Chain Monte Carlo (MCMC)-based approach to jointly analyze all fitted transit parameters and quantify their individual contributions to transit depth uncertainty through the structure of the full, multidimensional posterior. In this work, we address this gap and frame limb-darkening as an information bottleneck. Through injection--retrievals we quantitatively explore how transit depth precision and bias depend on the adopted limb-darkening prescription and prior assumptions. To ensure our conclusions are not tied to a particular stellar or wavelength regime, we \vtwo{inform} our framework with intensity profiles extracted from modern stellar atmospheric models spanning a wide range of $T_{\rm eff}$, $\log g$, $[{\rm M/H}]$, and $\lambda$. Building on this setup, we quantify the sensitivity of the transit depth to all fitted transit parameters, identify which correlations dominate information loss, and assess what complementary data are most effective at mitigating the TDPP.

\section{The transit-depth precision problem}\label{sec:TDPP}

The most informative observable for atmospheric science across the board is the transit depth, $D(\lambda) = (R_p(\lambda)/R_{\star}(\lambda))^2$, in which $R_p(\lambda)$ is the planetary radius, and $R_\star(\lambda)$ is the stellar radius. Measuring $D$ at different wavelengths reveals the planet's apparent size as a function of wavelength, which can be translated into information on atmospheric opacity \citep[e.g.,][]{Seager2000, Benneke2012, deWit2013}. The precision with which $D$ is measured directly sets the limit on the inferences that can be made about atmospheric characteristics \citep[e.g.,][]{Deming2009, Greene2016}. It is thus crucial that the uncertainty on the transit depth ($\sigma_D$) be minimized---as well as biases on it, naturally. Since variations in transit depth linked to atmospheric opacity are notably weaker than those linked to the TDPP, we assume hereafter that $D$ is wavelength-independent.

To first order, one could expect that $\sigma_D$ relates to the scatter in the photometric time series (\autoref{fig:fig1}, left panel), which includes both photon noise (roughly Gaussian at high count levels) and correlated noise (marginal to first order in many modern datasets), measurable as the out-of-transit scatter for a bin of duration similar to the transit duration ($\sigma_{\rm ref} = \sigma_{\rm OOT}/\sqrt{N}$ with $N$ in-transit points). Specifically, one expects that $\sigma_D = \mathbf{\sqrt{3/2}} \ \sigma_{\rm ref}$, as the typical out-of-transit baseline for transit observations is twice the duration of the transit event targeted (see \autoref{sec:appendix_sqrt} for a full derivation). However, when comparing these two quantities in many existing datasets, a significant discrepancy between expectation and reality emerges (\textit{i.e.}, $\sigma_D/\sigma_{\rm ref}\gg \mathbf{\sqrt{3/2}}$). 
To simplify notation in our subsequent investigations, we define the amplification factor $A = {\sigma_D}/\sigma_{\rm ref}$.

We illustrate this discrepancy through a set of injection--retrieval tests, whose results are shown in the right panel of \autoref{fig:fig1}. For these tests, we simulated a single transit with the \texttt{jaxoplanet} package \citep{jaxoplanet} for a hypothetical system (see \autoref{tab:tab1} for its properties) and varying levels of white noise, before fitting the mock data with the same transit model. We consider three families of tests, each using a different LDL for both injection and retrieval: a second-, third-, and fourth-order polynomial LDL, respectively. These three families are intended to illustrate how the dimensionality of the limb-darkening parameterization affects the TDPP, with each additional order introducing one more free parameter. The transit is constructed with equal pre-, post- and in-transit durations (as in \autoref{fig:fig1}, left panel). To sample the posterior, we used the affine\vtwo{-}invariant MCMC method implemented in the \texttt{emcee-jax} package \citep[a JAX implementation of \texttt{emcee},][]{emcee} with 50 walkers run for 100,000 steps, discarding the first 70,000 as burn-in. We applied uninformative uniform priors to all fitted parameters except the orbital period, on which we set a strong Gaussian prior, and the transit midpoint, which was fixed. This setup reflects the fact that such analyses generally assume prior observations have tightly constrained transit timings. Walkers were initialized in a wide multidimensional Gaussian ball around truth. Poorly converged samples are excluded via three rounds of iterative interquartile range (IQR) clipping\footnote{Clipping is applied to individual (walker, step) pairs rather than to individual chains. At each round, pairs whose $\chi^2$ or parameter values fall outside $\mathrm{median} \pm \{5, 4, 3\} \times \mathrm{IQR}$ are removed. \vtwo{Across the retrievals presented in this work this removes at most $5\%$ of samples, so the posterior widths we report reflect the sampled distribution rather than the clipping thresholds.}}. From $D$'s posterior, we measured $\sigma_D$, and with the injected white noise level, we computed the corresponding $A$. All details regarding the retrieval setup, including the LDC values used for injection in each of the three cases, can be found in \autoref{tab:tab1}.

\begin{figure}[tbp]
    \script{Fig2_plot.py}
    \includegraphics[trim = 0 05mm 0 0 ,width=1.05\linewidth]{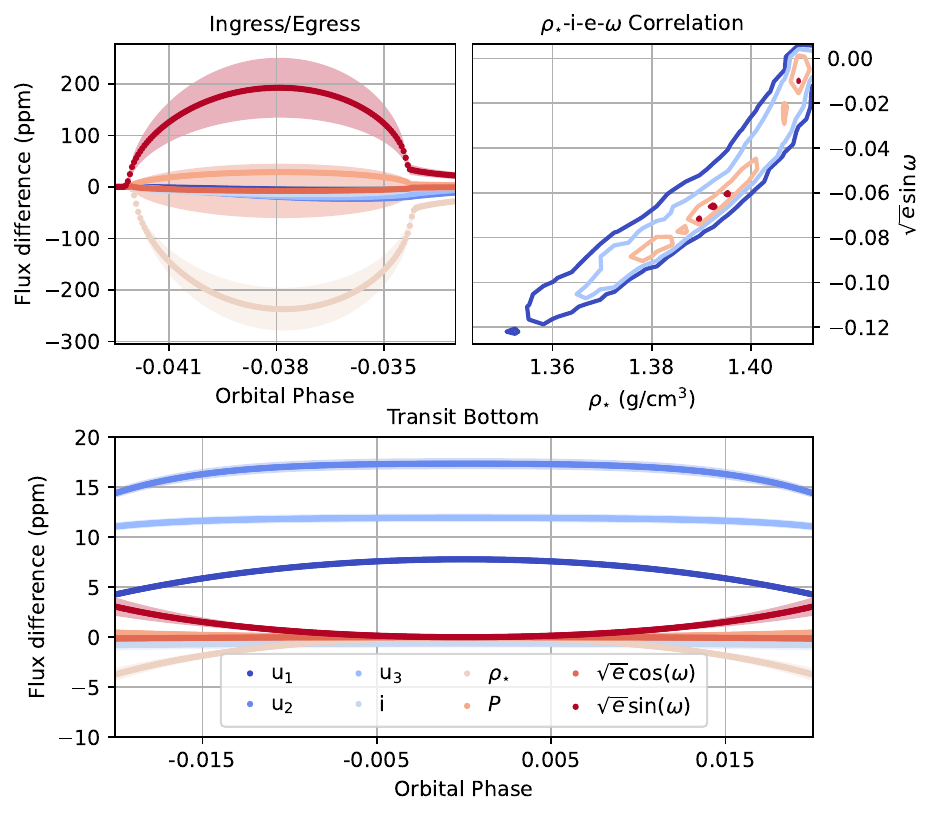}
    \caption{\textbf{Parameter-induced perturbations on transit shape.} Separate panels highlight the transit ingress/egress, and bottom. \vtwo{Markers and shaded regions correspond to the mean and standard deviation across 10 MCMC retrievals with different noise realizations.} Symmetric ``bumps'' can be clearly seen at ingress and egress, driven by the orbital parameters. In contrast, fluctuations in the bottom are dominated by the LDCs. Perturbations beyond the photometric precision (16~ppm here) are due to underlying correlation, as exemplified in the top-right panel by the anti-correlation between $\rho_{\star}$ and $\sqrt{e}\sin{\omega}$. Colored contours indicate the $10, 30, 60$, and $90\%$ iso-probability levels. In practice, such correlations couple all orbital parameters simultaneously.}
    \label{fig:fig2}
\end{figure}

Each colored box plot set in the right panel of \autoref{fig:fig1} highlights one of the injection--retrieval families. The behavior of these sets falls into four regimes. At high scatter, the amplification factor converges to the theoretical limit of \vtwo{$\sqrt{3/2}$}, indicating that the retrievals are noise-limited. As scatter decreases, a transition region is crossed wherein retrievals shift from being noise- to model-limited (shown as a shaded band). In the model-limited regime, $A$ plateaus at a value set by the degeneracies and correlations among the fitted model parameters. Indeed, when fitting transits, parameters describing the exoplanet's orbit and the stellar intensity profile must be inferred alongside the transit depth. Many of these affect the transit shape in similar ways, producing compensatory effects where different parameter combinations yield similar light curves \citep[see, e.g., \autoref{fig:fig2}; see also][their Figure~4]{deWit2012}. Such dependence broadens the posterior distribution of $D$, increasing $A$. The plateau value depends on the dimensionality of the retrieval: as more model parameters are included, for example when moving from the second- to fourth-order polynomial LDL, the space of iso-probability solutions expands, thus broadening the posterior. Finally, at extremely low scatter, the likelihood landscape becomes so sharply peaked that the sampler struggles to explore it efficiently, and $A$ is no longer limited by the model but by the ability of the MCMC to draw independent samples. This sampler-limited regime is indicated by the upward fade of the asymptote lines toward the left.

\section{An informed injection--retrieval framework}\label{sec:framework}
With the existence of the TDPP established (which holds regardless of the specific system or LDC values used) we now turn to diagnosing it quantitatively. Before doing so, we must specify the injection--retrieval framework underlying the subsequent analyses. As \autoref{sec:TDPP} makes clear, limb-darkening is at the core of the TDPP, thus the related choices we make directly shape what our analyses can reveal.

In particular, two choices are required: the LDL used for injection and retrieval, and the LDCs used for injection. Because our test system is hypothetical, no literature LDCs exist and making arbitrary choices could limit the generalizability of our findings. We therefore \vtwo{inform} both choices---particularly the latter---in modern stellar model grids, ensuring that our adopted setup spans representative regions of the relevant physical space.

We note that all injection--retrievals throughout this work assume a central-transit geometry ($b=0$). Grazing and near-grazing configurations introduce qualitatively different parameter degeneracies, and dedicated follow-up work varying the orbital properties of the hypothetical system---including its impact parameter---will be performed to determine how our recommendations generalize to those regimes.

\subsection{Selecting an appropriate law}\label{sec:appropriate_law}
All light curves generated for this section's analyses were produced using the fourth-order nonlinear LDL
\begin{equation}
    \frac{I(\mu)}{I(1)} = 1 - \sum_{k=1}^{4} c_k(1-\mu^{k/2})
\end{equation}
with $I(\mu)$ being the intensity at a given position on the stellar disk. This prescription has been shown to produce the most accurate representation of stellar intensity profiles and currently appears to be the most appropriate choice for high-precision transit observations (e.g., \citetalias{Espinoza2016}; \citealt{Morello2017, Keers2024}). For the retrieval law, we adopt a third-order polynomial LDL:
\begin{equation}
    \frac{I(\mu)}{I(1)} = 1 - \sum_{k=1}^{3} u_k(1-\mu)^k.
\end{equation}
We demonstrate in \autoref{sec:boosting_science} this provides the best compromise between flexibility, bias, and amplification factor at the noise level considered. Hereafter, $(c_1, c_2, c_3, c_4)$ refers to the coefficients of the fourth-order nonlinear LDL and $(u_1, u_2, u_3)$ to those of the third-order polynomial LDL.

\subsection{Selecting appropriate coefficients}\label{sec:appropriate_coeff}
Rather than selecting an arbitrary set of LDCs, we extract representative vectors $(c_1, c_2, c_3, c_4)$ from the population of physically plausible intensity profiles spanning the full range of stellar parameters and wavelengths available in stellar models. This ensures both the physical fidelity of our injection and a means of assessing how sensitive our conclusions are to the physical inputs.

\textit{Profile generation.} Using \texttt{ExoTiC-LD} \citep{ExoTiC-LD}, we extract stellar intensity profiles from the MPS-ATLAS set1 stellar models \citep{mpsset1} on a $10\times10\times10$ grid, uniformly sampling $T_{\rm eff} \in [3500, 9000]$\,K, $\log g \in [3.0, 5.0]$, and $[\rm M/\rm H] \in[-5.0, 1.5]$. Profiles are available at 1221 wavelengths ($\lambda \in [0.0099, 160]$\,$\mu$m) per grid point, yielding a total dataset size of $1{,}221{,}000$. We adopt MPS-ATLAS set1 as it is among the most up-to-date \vtwo{sets} of stellar models and yields LDCs consistent with its set2 counterpart \citep{mpsset2}. Restricting ourselves to only MPS-ATLAS set1 avoids introducing inter-model variability unrelated to our driving question. This choice also defines the parameter ranges explored, as different models cover the ($T_{\rm eff}$, $\log g$, $[\rm M/\rm H]$, $\lambda$)-space \vtwo{differently}. 

\vtwo{We note, however, that this choice is not without consequence: 1D models, such as MPS-ATLAS, predict systematically different LDCs than their 3D counterparts \citep[e.g.,][]{Hayek2012, Magic2015}. While the literature is not fully consistent on the direction of this discrepancy, there is broad agreement that differences in the underlying physics and geometry between 1D and 3D models meaningfully impact the resulting LDCs. Our results are therefore conditioned on the MPS-ATLAS model assumptions: the specific values of $A$ and transit depth biases we report should be interpreted within this 1D framework, rather than as an absolute characterization with real stellar intensity profiles. Repeating this analysis with different stellar model grids, and in particular with 3D models, is an important direction for future work but lies outside the scope of this study.}

\begin{figure*}[ht]
    \script{Fig3_plot.py}
    \centering
    \includegraphics[trim = 0 05mm 0 0 , width=0.95\textwidth]{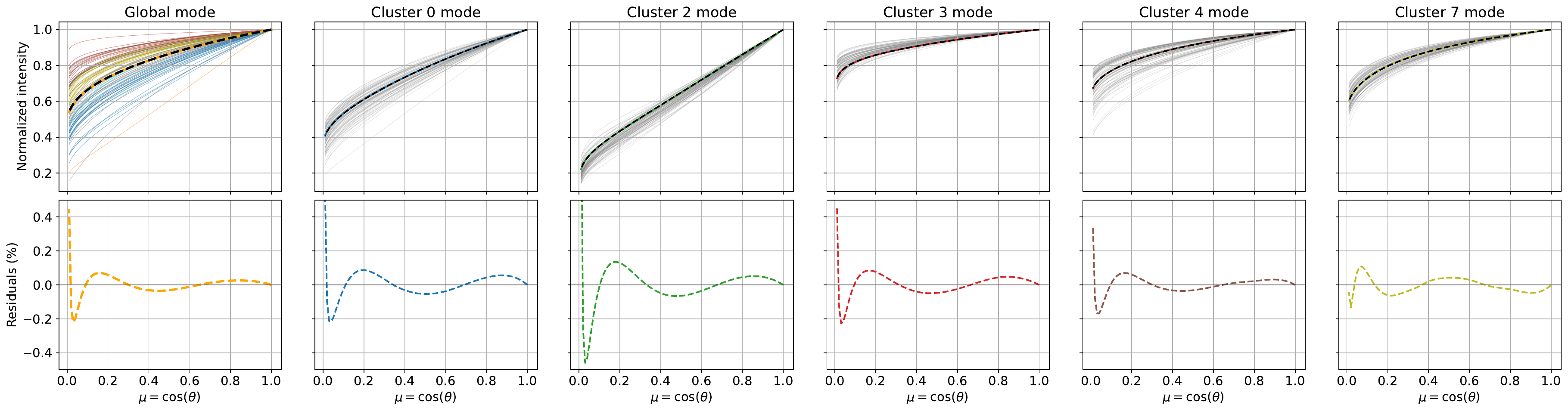}
    \caption{\textbf{Representative intensity profiles.} Intensity profiles for the global and cluster modes introduced in \autoref{sec:framework} and used in the analyses of \autoref{sec:amp_factor_cause} and \autoref{sec:boosting_science}. \emph{Top:} A random subset of 100 profiles is shown in each panel's background: for the global mode panel, profiles are colored by their cluster assignment (see \autoref{fig:appendix_fig3}); for each cluster mode panel, profiles belonging to that cluster are shown in gray. The modal intensity profile for each panel is overlaid as a solid colored line, with its fourth-order nonlinear LDL fit shown as a dashed black line. The resulting LDCs along with the stellar parameters and wavelength used to generate each profile are reported in \autoref{tab:tab1}. \emph{Bottom:} Relative difference residuals between the fourth-order nonlinear LDL fit and the true intensity profile.}
    \label{fig:fig3}
\end{figure*}

\textit{Filtering.} We restrict the wavelength coverage to the JWST NIRSpec PRISM range ($0.6$--$5.3$\,$\mu$m) for two reasons: it targets a broad wavelength region relevant for most JWST transit observations, and it removes the grid boundaries where numerical artifacts produce nonmonotonic intensity profiles\footnote{At extreme wavelengths, intensity values become very small causing numerical precision losses that allow intensities at adjacent $\mu$ values to overlap spuriously. For instance, $I(\mu=0)\gt I(\mu=0.25)$, which is nonphysical since intensity decreases from the disk center outwards.}. After this cut, $585{,}000$ profiles remain. We then manually mask any residual nonmonotonic profiles, removing an additional 441 (${\lt}0.1\%$). These preferentially originate from high wavelengths ($2.52$--$5.29$\,$\mu$m) and from low $T_{\rm eff}$, $\log g$, and $[\rm M/\rm H]$---regimes where limb-darkening is intrinsically weak, either because the plane-parallel approximation breaks down (low $\log g$) or because atmospheric complexity increases (e.g., molecular opacities at low $T_{\rm eff}$, or H$^-$ continuum dominating at low $[\rm M/H]$). In both cases, the center-to-limb intensity contrast is sufficiently reduced that numerical noise can produce spurious profile inversions. A detailed examination of these artifacts and their dependence on stellar parameters is left to future work. Our final dataset comprises $584{,}559$ profiles.

\textit{Coefficient extraction.} Each profile is interpolated onto a uniform 100-point $\mu$-grid \vtwo{spanning $[0.01, 1.0]$} via a cubic spline and fit with the fourth-order nonlinear LDL\footnote{Throughout this work, all LDL fits to an intensity profile are performed via nonlinear least-squares optimization using \texttt{lmfit} \citep{lmfit}.}. This approach to intensity profile fitting, first presented by \citet{Claret2011}, was shown by \citet[][hereafter \citetalias{Espinoza2015}]{Espinoza2015} to minimize the residuals between the true profile and its parametric approximation. 
\vtwo{By fitting down to $\mu = 0.01$, we include intensity points very close to the stellar limb, where the nonlinearity of intensity profiles is strongest and where 1D and 3D stellar models diverge most. As a result, the inferred LDCs may reflect nonlinear behavior specific to the 1D models' approximation rather than genuine limb-darkening. We assess how much this choice matters in \autoref{sec:compare_empirics} and \autoref{sec:appendix_mu_min}.
This LDC extraction procedure reduces} the dataset from a ($T_{\rm eff}$, $\log g$, $[\rm M/\rm H]$, $\lambda$)-dependent intensity field to a 4D space of LDC vectors $(c_1, c_2, c_3, c_4)$. The resulting distribution is shown in \autoref{fig:appendix_fig1} and \autoref{fig:appendix_fig2}, where the clear structure reflects the inherited dependence on the underlying physical parameters. A detailed exploration of this topology is left to future work.

\vtwo{\textit{Subset selection.}
To reduce the computational strain on the following steps, we extract a subset from the available $584{,}559$ vectors. Our selection follows two criteria: first, the subset must be large enough to preserve the overall structure of the 4D LDC space and small enough for subsequent computations to be tractable; second, the subset must be representative of the population of transiting exoplanet-hosting stars. We determine empirically that the largest subset satisfying the first criterion is $50{,}000$. 
To satisfy the second, we extract the 3D distribution of ($T_{\rm eff}$, $\log g$, $[\rm M/\rm H]$) values for all stars known to host transiting exoplanets from the NASA Exoplanet Archive\footnote{\url{https://exoplanetarchive.ipac.caltech.edu}, queried on July 8$^{\rm th}$ 2026.} and fit a 3D kernel density estimator (KDE) to this distribution. We use the resulting KDE to perform an importance-weighted draw of $50{,}000$ vectors from the full dataset.}

\textit{Clustering and mode identification.} To extract representative LDC vectors from this dataset, we apply hierarchical agglomerative clustering \citep{Everitt2011} \vtwo{to the extracted subset of vectors}. We used the Mahalanobis metric with Ward linkage \citep{Ward1963} and a distance cutoff of $80$, all selected empirically by visually inspecting the resulting clustering across a range of combinations. Singleton clusters are reassigned to the nearest non-singleton cluster. Our approach yields nine clusters, whose membership is shown in \autoref{fig:appendix_fig3}. For each of them, we identify the LDC vector closest to its centroid as the ``cluster mode.'' The vector closest to the global centroid of all points is designated as the ``global mode,'' representing the single most typical profile across the physical space explored.
The global mode is used as the injection vector for the analyses in \autoref{sec:amp_factor_cause}. In \autoref{sec:boosting_science}, we inject with the modes of \vtwo{five clusters (clusters 0, 2, 3, 4, and 7}; see \autoref{fig:fig3}), testing how our conclusions generalize across the ($T_{\rm eff}$, $\log g$, $[\rm M/\rm H]$, $\lambda$)-space. \vtwo{The global mode and five selected cluster modes span representative regions of the explored LDC space, sampling a range of stellar types, metallicities, and observing wavelengths. Their LDC vectors and corresponding physical parameters are reported in \autoref{tab:tab1}.}

\section{Diagnosing the TDPP}\label{sec:amp_factor_cause}

Equipped with an injection--retrieval framework, we now investigate what drives the increase in $A$ beyond the noise-limited regime ($\sigma_{\rm OOT} \lesssim 30{,}000$\,ppm). As \autoref{sec:TDPP} makes clear, this amplification originates from degeneracies between model parameters. Here we identify which parameters are responsible and quantify their relative contributions.

\begin{figure*}[ht!]
    \script{Fig4_plot.py}
    \centering
    \includegraphics[width=1\textwidth]{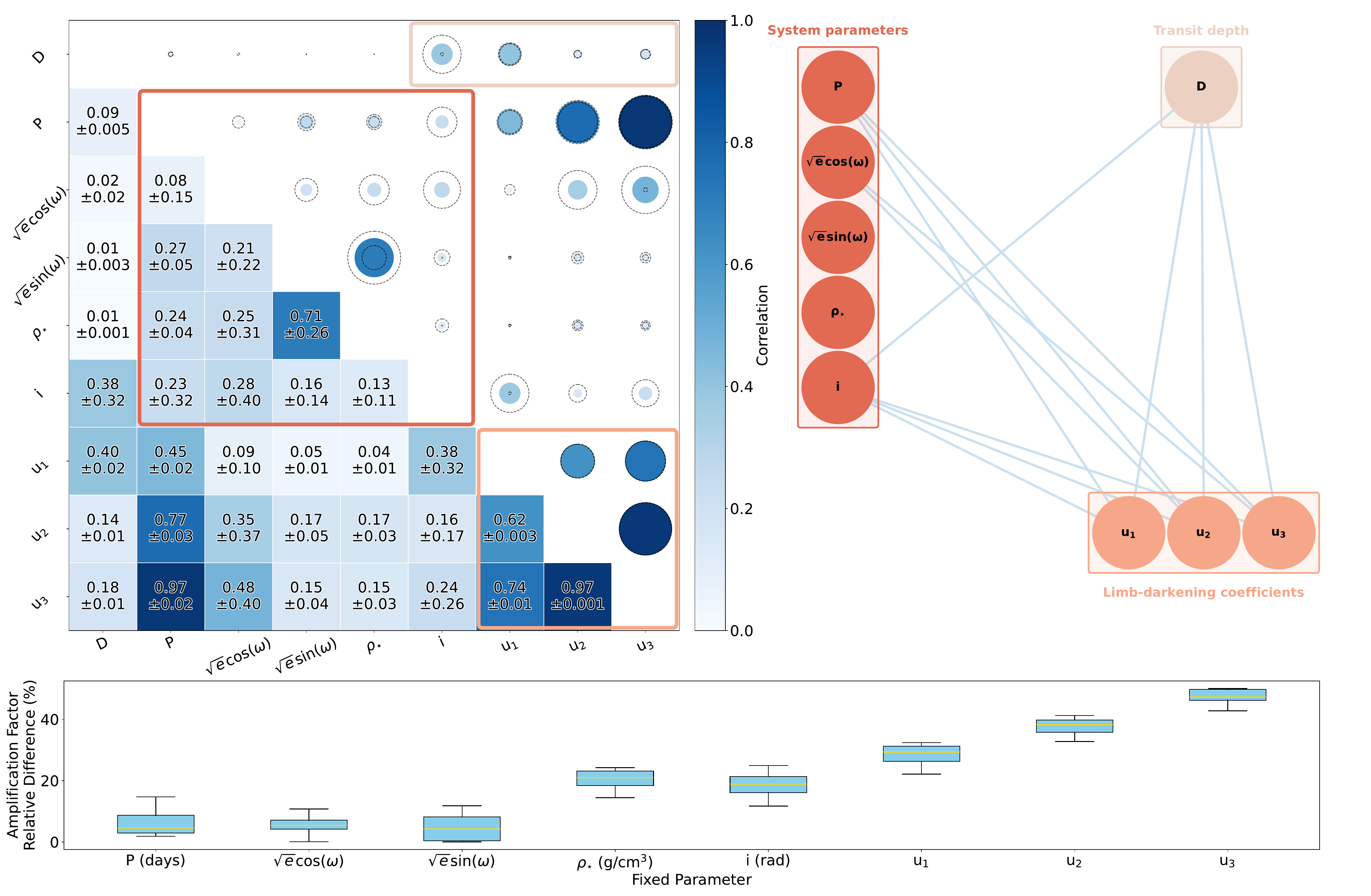}
    \caption{\textbf{Revealing the underlying network of correlations.} \textit{Left:} Correlation matrix resulting from the sensitivity analysis. \vtwo{The lower triangle reports the mean and standard deviation of correlation values derived from ten $\chi^2$ maps produced from different noise realizations. Shading is scaled to the absolute values of the means. Colored boxes highlight families of related parameters, which are easier to distinguish in the upper triangle's circular representation of the same matrix. There, disk size and shading scale with correlation strength, while a pair of dashed circles marks the $\pm1\sigma$ spread.} The transit depth is most strongly correlated with the polynomial LDCs $u_1$, $u_2$, and $u_3$\vtwo{, followed by the inclination $i$}. \textit{Right:} Network representation of the same correlations with parameters grouped into families using the same color scheme as in the matrix. Blue lines highlight the strongest links between parameters of different families. \vtwo{\textit{Bottom:} Relative difference in the amplification factor between fits in which the parameter on the $x$-axis is fixed to its true value and fits in which it is left free. Each box plot summarizes ten injection--retrievals with different noise realizations, following the same convention as \autoref{fig:fig1}. The LDCs produce the largest changes in $A$, followed by $i$ and $\rho_\star$, while $P$, $\sqrt{e}\cos{\omega}$, and $\sqrt{e}\sin{\omega}$ leave $A$ essentially unchanged.}}
    \label{fig:fig4}
\end{figure*}

\subsection{Perturbation analysis}\label{sec:perturbation_analysis}

The model-driven increase in $A$ results from correlations between transit model parameters. Different combinations can generate nearly identical light curves, making it difficult to disentangle each parameter's contribution and recover its true value. A simple way to illustrate this effect is to perturb each parameter by 1$\sigma$ (measured from the posterior output by \vtwo{an} MCMC) and examine how the light curve responds, an approach previously used by \citet{deWit2012} to study secondary eclipses. \vtwo{To capture the spread of these perturbations, we repeat this procedure across 10 MCMC retrievals with different noise realizations and, for each parameter, compute the mean and standard deviation of the perturbation curves across seeds.} We present the results of this analysis in \autoref{fig:fig2}, specifically focusing on changes in the transit ingress/egress (top left panel), and center (bottom panel).

At ingress and egress, perturbations generate symmetric bumps, with orbital parameters producing the largest changes. This is expected: fixing the transit midpoint and varying the orbital parameters is equivalent to altering the transit duration \citep{Seager2003}. Parameters that induce similar amplitude bumps but of opposite signs, like the stellar density ($\rho_{\star}$) and $\sqrt{e}\sin{\omega}$, are anti-correlated (illustrated in \autoref{fig:fig2} top right panel). These relationships complicate posterior sampling, as algorithms like MCMC or nested sampling must navigate a landscape filled with local minima. In contrast, LDCs have little to no impact on the transit wings; variations in the LDCs impact the intensity profile but not the transit duration. 

The picture is different in the transit center/bottom. Orbital parameters induce modest changes ($\sim$0--3\,ppm), whereas perturbations to the LDCs shift the bottom by $\sim$5--15\,ppm. This reflects the role of limb-darkening in shaping the overall depth and curvature of the transit bottom. The three coefficients produce similar patterns, differing mostly in amplitude. Since $D$ is measured from the transit bottom, these results reinforce the conclusion that LDCs are the main drivers of the TDPP.

\subsection{Sensitivity analysis}\label{sec:sensitivity_analysis}

To quantify parameter correlations and their impact on $D$, we carried out a targeted sensitivity analysis. The challenge here is to isolate correlations between pairs of parameters without contamination from the rest of the parameter space. This cannot be done via cuts in the full multidimensional posterior sampled by an MCMC, as marginalizing over the other parameters introduces biases. Instead, we evaluate the local $\chi^2$ (or log-likelihood) surface over two chosen parameters at a time, holding all others fixed to truth.

For each pair, we generated a light curve, injected 16\,ppm of white noise, placing the retrievals in the model-limited regime (see \autoref{fig:fig1}), and computed the $\chi^2$ across a grid of values centered on truth. Because our injection and retrieval LDLs are different (see \autoref{sec:appropriate_law}), we define ``truth'' in this case, as the LDC vector $(u_1, u_2, u_3)$ obtained from fitting the injected fourth-order nonlinear LDL with the retrieval's third-order polynomial LDL.

Elliptical iso-$\chi^2$ contours were then overlaid \vtwo{on} each map. The ellipses' rotation angle $\theta$ encodes the space's orientation, which we translate into a correlation metric $c = \mathrm{sin}(2\theta)$. $c$ peaks for diagonal contours (\textit{i.e.}, strong correlation) and reaches a minimum for vertical or horizontal contours (\textit{i.e.}, no correlation). 

The results are summarized in \autoref{fig:fig4}, and the full set of $\chi^2$ maps is provided in \autoref{fig:appendix_fig4}. The left panel presents the correlation matrix. \vtwo{As in the perturbation analysis, we repeat the correlation-coefficient extraction across 10 noise realizations so that each entry in the matrix reports a mean and a $\pm1\sigma$ spread.} From this matrix, \vtwo{three} conclusions can be drawn. First, \vtwo{the LDCs are correlated with $D$, in line with the perturbation analysis and the results from \autoref{fig:fig1}, confirming them as strong contributors to the TDPP}. \vtwo{Second, $i$ also correlates significantly with $D$---unexpectedly so, since \autoref{fig:fig2} shows it to be the only orbital parameter that leaves the transit ingress, egress, and bottom unperturbed.} \vtwo{Finally,} two intraconnected families of correlations can be distinguished: one for orbital parameters and another for the LDCs. However, these families are not isolated: their members are clearly linked, most importantly $\sqrt{e}\cos{\omega}$, $i$, $P$ and $u_2$, $u_3$---corresponding to the ``$e$-$b$-$\rho$-BD” (for brightness distribution) coupling introduced in \citet{deWit2012}. This mirrors the results of \autoref{fig:fig2}, where correlated parameters (either within each family or across them) produced similar perturbations in transit shape. \vtwo{The spread in correlation values is parameter-dependent: $\sqrt{e}\cos{\omega}$ and $i$ vary far more across realizations than $u_1$ and $P$.} Overall, the LDCs \vtwo{and the inclination} affect $D$ through two channels: directly, via their own strong correlation with $D$, and indirectly, by serving as a conduit through which orbital parameters propagate their uncertainty into $D$.

\vtwo{To assess how these couplings translate into $A$, we ran a separate set of injection--retrieval tests in which each parameter was, in turn, fixed to its true value. This complements the sensitivity analysis' pairwise view with a multidimensional approach, and measures how much $A$ would improve were each parameter known perfectly. We stress that these are individual sensitivities rather than a decomposition of $A$ into independent contributions: because the fitted parameters are mutually correlated, fixing one also suppresses the degeneracy pathways it participates in, so the resulting fractional changes are not additive. The bottom panel of \autoref{fig:fig4} reports the results, with each box plot summarizing the relative difference in $A$ between the fixed- and free-parameter fits across 10 noise seeds. The LDCs dominate, from $\sim30\%$ ($u_1$) to $\sim50\%$ ($u_3$), consistent with the strong influence on $D$ established above. $i$ and $\rho_\star$ follow---expected for $i$ given the sensitivity analysis results, but not for $\rho_\star$. We attribute the $\sim25\%$ shift from fixing $\rho_\star$ to the ``$e$-$b$-$\rho$-BD" pathway: its coupling to the other fitted parameters helps channel uncertainty into $D$, inflating its posterior and hence $A$. Mirroring the correlation matrix, $P$, $\sqrt{e}\cos{\omega}$, and $\sqrt{e}\sin{\omega}$ leave $A$ essentially unchanged.}

\vtwo{Together these three analyses paint a consistent picture: the LDCs, and to a lesser extent $i$, are at the heart of the TDPP, driving the increase in $A$ both directly through their strong correlation with $D$ and indirectly through the orbital parameters to which they couple. This underscores the need for tighter external constraints: on LDCs, through improved stellar models, and on the orbital parameters, through complementary observations such as spectroscopic time series.}

\section{Reliably boosting transit science}\label{sec:boosting_science}

\begin{figure*}[htbp]
    \script{Fig5_plot.py}
    \centering
    \includegraphics[width=0.85\textwidth]{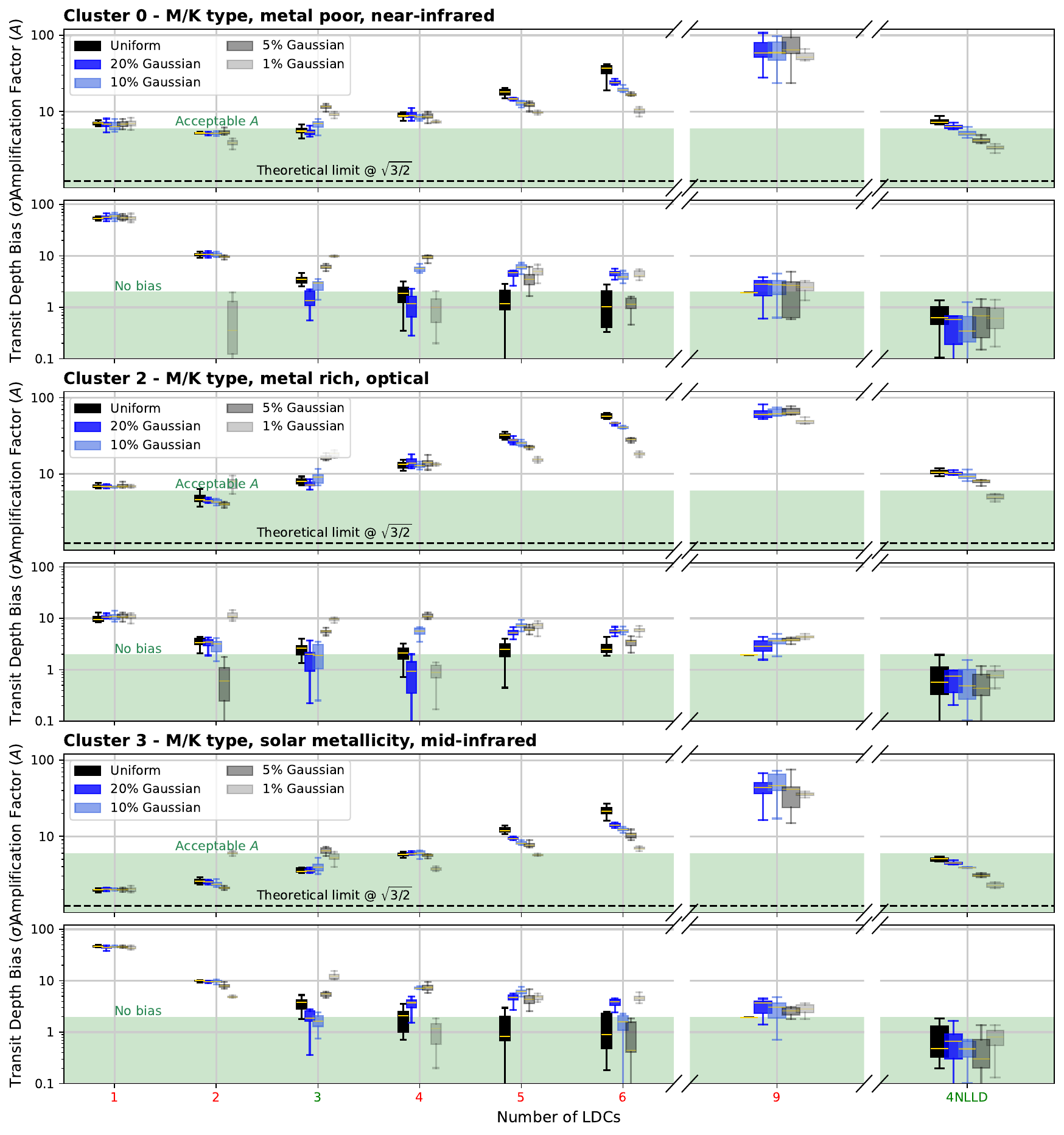}
    \caption{\textbf{Toward optimal limb-darkening \vtwo{parameterization}}. Amplification factor (top panels) and transit depth bias (bottom panels) from our injection--retrieval tests shown for \vtwo{five} representative cluster modes. Results are shown as a function of the number of LDCs (x-axis) under different limb-darkening \vtwo{prior} assumptions, ranging from uniform priors to Gaussian priors of varying widths. Polynomial LDLs are used for cases with 1-9 LDCs, while ``4NLLD" refers to the fourth-order nonlinear LDL. Since this is the injection LDL, it yields the best performance.
    Priors informed by current stellar models are highlighted in blue ($10\%$--$20\%$). As in \autoref{fig:fig1}, the black dashed horizontal line marks $A$'s theoretical limit, and box plots summarize the results from 10 MCMC retrievals with different noise realizations. Green vertical bands indicate the regimes considered acceptable: for the bias, results within $2\sigma$ of truth and for the amplification factor, $A \leq 6$.  As discussed in \autoref{sec:impact_prior_choice}, parameterizations meeting both criteria are highlighted with green x-axis labels, and those failing one or both are in red. \vtwo{Clusters 0, 2, and 3 are shown here; clusters 4 and 7 are continued in \autoref{fig:appendix_fig6}.}}
    \label{fig:fig5}
\end{figure*}

\vtwo{The LDCs are the primary drivers of the TDPP, with the orbital parameters---mainly the inclination---playing a secondary role. In this section we discuss mitigation strategies for constraining the LDCs and quantify the gains in both $A$ and transit depth bias that such strategies could deliver, before turning to the orbital parameters.}

\subsection{Constraints from stellar models}\label{sec:constraint_stellar_model}

A common practice when fitting transits is to constrain LDCs using model-based estimates. Tools such as \texttt{ExoTiC-LD} or \texttt{ExoCTK} \citep{ExoCTK} provide LDC estimates from a set of stellar parameters and a given stellar atmospheric model. While these estimates are convenient, it is important to remain cautious about biases introduced when relying on them in transit fits. Indeed, \citetalias{Espinoza2015} and \citetalias{Espinoza2016} showed that LDCs vary significantly between stellar models and \citet{Howarth2011} further warned that model-derived LDCs fundamentally differ from those measured from photometric observations. Using model-derived LDCs without caution can therefore introduce systematic biases that depend on the stellar type and orbital configuration. \citetalias{Espinoza2015} and \citetalias{Espinoza2016} made two additional points: LDCs should not be fixed to model values but allowed to float during fitting; and some models, such as ATLAS9 \citep{Kurucz1979}, better reproduce photometric limb-darkening than others, like PHOENIX ACES \citep{Husser2013}. While these studies caution against the blind use of model-derived LDCs, they also show that such outputs still provide good order-of-magnitude estimates \vtwo{of} the values expected from photometry (see Figures 8 \& 9 of \citetalias{Espinoza2015}).

Motivated by these insights, we test how varying the number of free LDCs and the strength of priors imposed on them affects both the amplification factor and the transit depth bias. To set realistic priors, we estimate the spread in LDCs between the MPS-ATLAS set1 and set2 models \citep{mpsset1,mpsset2}. The set1 LDC distribution was already derived in \autoref{sec:appropriate_coeff}; we therefore generate a parallel set2 distribution using identical $T_{\rm eff}$, $\log g$, and $[{\rm M/H}]$ grids, and wavelength-filtering steps. The two distributions are compared point-by-point by matching profiles at identical physical parameters, and computing the signed relative difference $(c_{\rm set2} - c_{\rm set1})\,/\,|c_{\rm set1}|$ for each pair. The resulting distributions are summarized as histograms in the top panels of \autoref{fig:appendix_fig5}. The medians are near zero for all four coefficients ($\leq1.0\%$), confirming that the two sets are effectively unbiased relative to one another. The inter-model spread, however, grows with coefficient order: the IQR is $5\%$ for $c_1$, $15\%$ for $c_2$, and $18\%$--$19\%$ for $c_3$ and $c_4$. This trend is expected, as higher-order coefficients capture increasingly fine-scale variations in the intensity profile which are, by definition, weaker and harder to constrain. We therefore adopt the IQR of the higher-order coefficients ($\sim$10--20\%) as the tightest justifiable Gaussian prior width that can be placed on LDCs from current stellar models. We note, however, that the distributions exhibit long, asymmetric tails, and the bottom panels of \autoref{fig:appendix_fig5} show that spread grows substantially at extreme stellar parameters---particularly at $[{\rm M/H}] \lesssim -3$, $T_{\rm eff} \gtrsim 5500$\,K, and short wavelengths. The adopted prior widths should therefore be seen as summary statistics, with inter-model differences potentially shrinking or growing substantially depending on the physical regime considered.

We restrict this comparison to the MPS-ATLAS family, because the two sets share the same underlying atmospheric assumptions\footnote{Differing in their elemental abundances and convection modeling.} and span identical grids of stellar parameters and wavelengths, simplifying comparison. Other widely used models (e.g., PHOENIX, \citealt{Husser2013}; Stagger, \citealt{Magic2013}; Kurucz, \citealt{Kurucz1979}) are deliberately excluded as their non-overlapping parameter coverage and distinct underlying physics would conflate genuine physical differences with model specification artifacts. Nevertheless, a full sensitivity analysis comparing all available models to photometric measurements, akin to \citetalias{Espinoza2015}, remains necessary to fully validate the LDCs derived from this model set. 

\subsection{Impact of priors and LDL choice}\label{sec:impact_prior_choice}

\autoref{fig:fig5} \vtwo{and \autoref{fig:appendix_fig6} present our injection--retrieval results in five panels, one per cluster mode used as the injection LDC vector (clusters 0, 2, 3, 4, and 7} in \autoref{fig:fig3}; see \autoref{sec:appropriate_coeff}). The mock light curves were generated with a fourth-order nonlinear LDL (as discussed in \autoref{sec:framework}) and injected with 16\,ppm of Gaussian white noise. Each panel sweeps over Gaussian priors of relative width\footnote{Relative meaning that for a LDC with true value $0.1$, a width of $20\%$ translates to a Gaussian prior with mean $0.1$ and standard deviation $\sigma = 0.2\times0.1=0.02$.} $f \in \{20, 10, 5, 1\}\%$ alongside an uninformative wide uniform prior, and over a broad range of retrieval LDLs: second-, third-, fourth-, fifth-, sixth-, and ninth-order LDLs, as well as \vtwo{a} fourth-order nonlinear LDL. Because the injection and retrieval LDLs generally differ, the prior means cannot simply be set to the injected coefficients. Instead, we fit the profile produced by the injected fourth-order nonlinear LDL with whichever retrieval LDL is under consideration and use the resulting LDCs as the Gaussian prior means. The \vtwo{five} injection vectors are chosen to span distinct regions of the  ($T_{\rm eff}$, $\log g$, $[\rm M/H]$, $\lambda$)-space, allowing us to assess how robustly our findings generalize. All other aspects of the model setup, posterior sampling, and post-processing match \vtwo{those in} \autoref{sec:TDPP} and are summarized in \autoref{tab:tab1}. 

Five main conclusions emerge.
\vtwo{First, the often-used quadratic (second-order polynomial) LDL is inadequate. For nearly all prior widths tested, it results in substantial transit depth biases. The magnitude of the bias is set by how much of the injected intensity profile lies outside the quadratic basis. Concretely, the $c_2$ and $c_4$ terms of the injected fourth-order nonlinear LDL are exactly representable as quadratic combinations of $(1-\mu)$, but the half-integer-power terms $c_1(1-\mu^{1/2})$ and $c_3(1-\mu^{3/2})$ are not. No choice of quadratic LDCs can reproduce them, and it is this mismatch that the transit depth must absorb.}

\vtwo{This is why bias does not simply track the overall magnitude of LDCs. Each half-integer term is almost reproducible by a quadratic law, and the small part of each that is not shares a common shape---but with opposite sign between the two. Counter-intuitively, this means that with $c_1$ and $c_3$ of the same sign, the nonquadratic-LDL contributions of $(1-\mu^{1/2})$ and $(1-\mu^{3/2})$ can cancel, leaving a profile the quadratic law can describe well and thus a low transit depth bias. However, the cancellation is not one-to-one: $(1-\mu^{3/2})$ departs from the quadratic basis far less than $(1-\mu^{1/2})$ does, so $c_3$ must be larger than $c_1$ to have this cancellation.}

\vtwo{This interplay governs the biases of quadratic law retrievals in \autoref{fig:fig5} and \autoref{fig:appendix_fig6}. Cluster 7 is the only cluster in which $c_1$ and $c_3$ carry opposing signs. Their mismatches compound rather than cancel, yielding a profile poorly described by a quadratic LDL and the largest bias across clusters. Conversely, cluster 2's LDCs, despite having by far the largest values ($c_1 = 0.76$, $c_3 = 1.42$), have the same sign and roughly the ratio needed for near-complete cancellation, leading to the lowest bias.}

\vtwo{We conclude that reducing the number of LDCs is not a viable route to mitigating the TDPP: if the chosen LDL lacks the expressivity to capture the true profile, large biases are unavoidable. Importantly, the bias incurred from misspecifying the intensity profile is \emph{absorbed into $D$, not into the fitted quadratic LDCs themselves}, which is precisely why such issues cannot be diagnosed by checking LDC values against model predictions.}

\vtwo{This also explains an otherwise counter-intuitive feature of \autoref{fig:fig5}: for quadratic retrievals, tightening the priors from $5\%$ to $1\%$ leads to unexpected jumps in $A$ and bias. The prior means are, by construction, the quadratic LDCs that best reproduce the injected profile---but those are not the same LDCs that best recover $D$. At $5\%$ the distance between these two optima remains within reach of the sampler, but at $1\%$ the LDCs are effectively pinned to the prior mean, unable to move toward the other optimum. The orbital parameters must then take up the compensation, increasing their correlation with $D$. This translates to shifts in both bias and $A$, although the exact direction of these changes depends on the cluster considered.}

\begin{figure*}[htbp]
    \script{Fig6_plot.py}
    \centering
    \includegraphics[width=\textwidth]{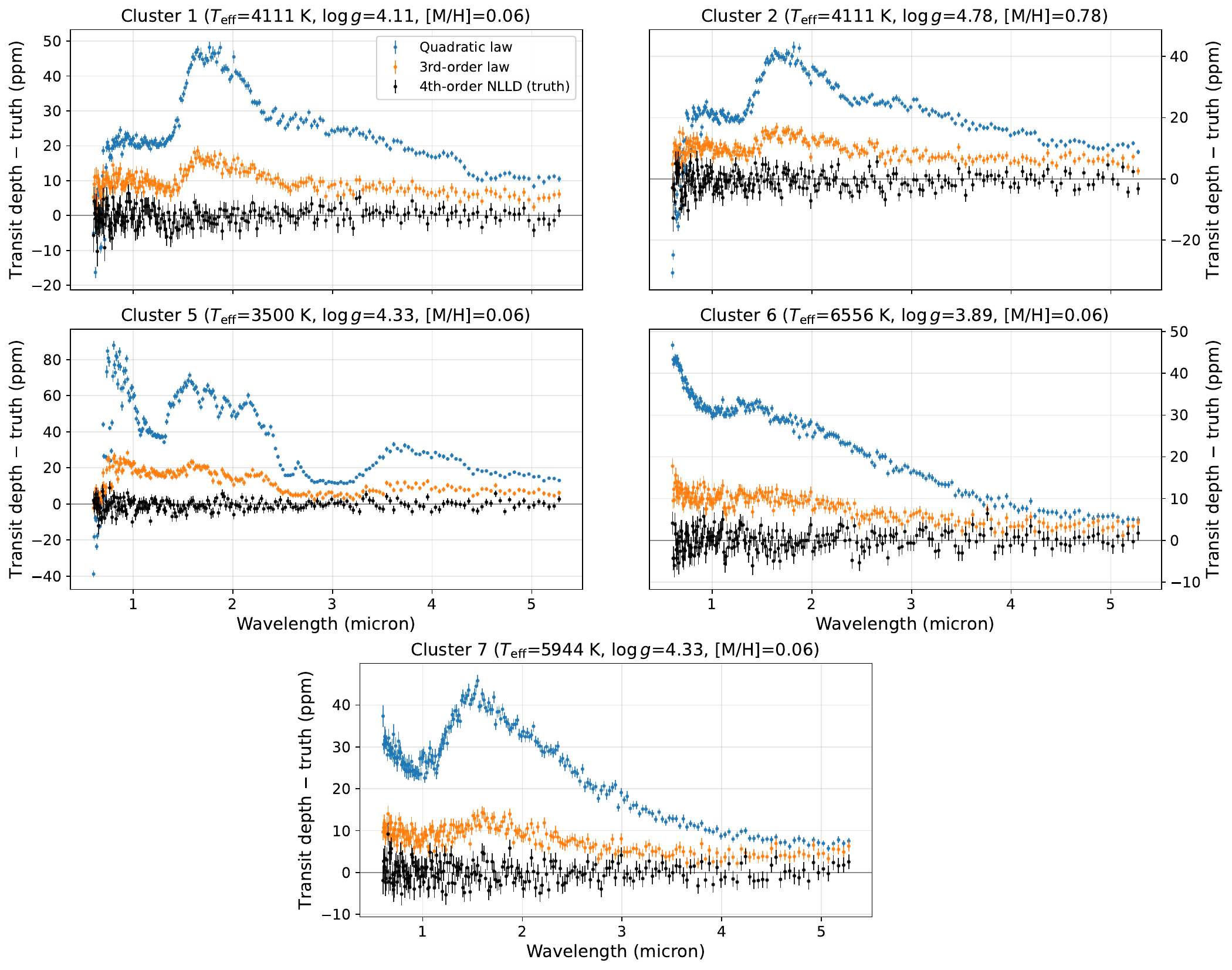}
    \caption{\vtwo{\textbf{Chromatic transit depth bias incurred by the choice of LDL.} Each panel corresponds to a set of stellar parameters, labeled above it, and shows the retrieved transit depth relative to the injected truth as a function of wavelength across the NIRSpec/PRISM range. For every wavelength, we generated a light curve with the fiducial system parameters of \autoref{tab:tab1} and the fourth-order nonlinear LDL coefficients obtained from a fit to the corresponding MPS-ATLAS set1 intensity profiles, then retrieve with three laws: the quadratic (blue), third-order polynomial (orange) and injection law (black). The injected transit depth is achromatic by construction, so any structure in the spectra originates solely from the limb-darkening prescription. Points and error bars give the median, $16^{\rm th}$, and $84^{\rm th}$ percentiles of the post-processed transit depth posterior. The corresponding amplification factors are shown in \autoref{fig:appendix_fig7}. Results are discussed in \autoref{sec:chrom_transit_depth_bias}.}}
    \label{fig:fig6}
\end{figure*}

\vtwo{Second, the adequacy of the quadratic law varies across the physical parameter space. The aforementioned cancellation condition is innately tied to where an intensity profile sits in LDC space and, by extension, in ($T_{\rm eff}$, $\log g$, $[{\rm M/H}]$, $\lambda$)-space (exemplified in \autoref{fig:appendix_fig1} and \autoref{fig:appendix_fig2}). Because $c_1$ and $c_3$ vary with all four, so does the transit depth bias incurred by a quadratic fit.
While we investigate wavelength dependence in \autoref{sec:chrom_transit_depth_bias}, mapping it across the full LDC space, and identifying which regions of $(T_{\rm eff}, \log g, [{\rm M/H}], \lambda)$ are most at risk of quadratic fits, is left to future work. The practical implication is that the suitability of a quadratic law is a property of the target and bandpass considered: a parameterization adequate for one star, or one wavelength bin, may be inadequate for another.}

Third, polynomials of third-order and higher result in more reliable retrievals, \vtwo{producing biases within the $2\sigma$--$3\sigma$ range for priors with $20\%$ width and below $2\sigma$ for $10\%$---both at the level currently achievable from stellar models.} The third-order polynomial LDL emerges as a practical ``sweet spot'': flexible enough to approximate the true intensity profile without introducing excessive degeneracy, yet not so high-dimensional that the amplification factor becomes unmanageable. At fourth order and beyond, the added degrees of freedom inflate $A$ rapidly without any compensating gain in accuracy: \vtwo{for cluster 2 under a uniform prior, $A$ reaches $\sim13$ at fourth order, $\sim56$ at sixth order, and exceeds $100$ at ninth order}---placing the corresponding box plot above the plotted range. This behavior is the expected consequence of over-\vtwo{parameterization}: with nine free LDCs and no informative priors, the higher-order polynomial LDCs are essentially unconstrained by the data, and the resulting LDC--$D$ degeneracies broaden $D$'s distribution.

\begin{figure*}[htbp]
    \script{Fig7_plot.py}
    \centering
    \includegraphics[width=\textwidth]{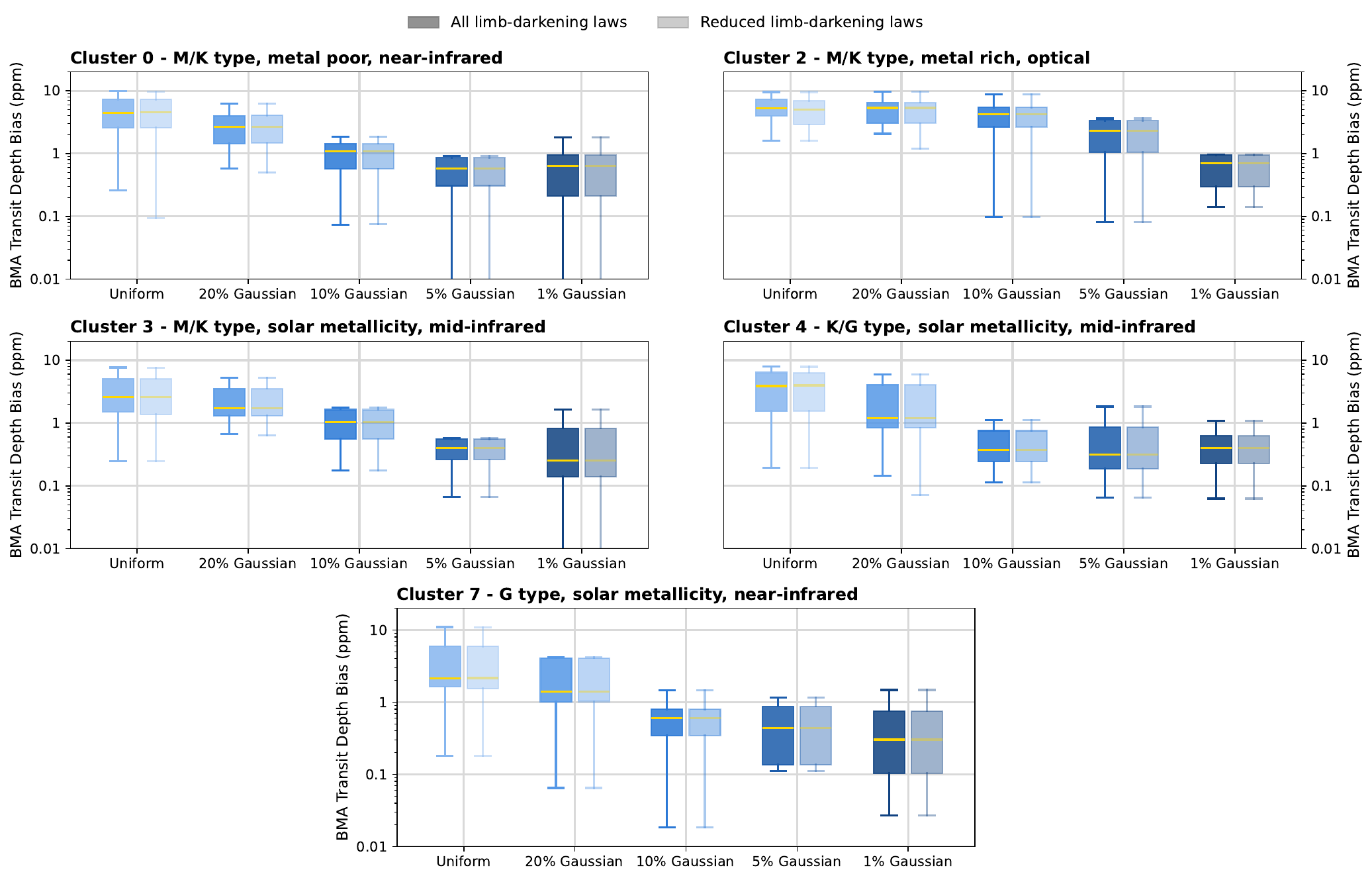}
    \caption{\vtwo{\textbf{Bayesian model average compared to injected transit depth}. Difference between the BMA transit depth and the injected truth, shown for each of the five cluster modes used as injection vectors, as a function of the prior strength imposed on the LDCs. \textit{Left column:} BMA computed from all LDLs considered in \autoref{fig:fig5}. \textit{Right column:} BMA computed from a restricted set of laws (linear, quadratic, third-order polynomial and fourth-order nonlinear), closest to the four used in \citetalias{Sing2026}. Box plots summarize results across 10 MCMC retrievals with different noise realizations. The BMA is within 10 ppm of the injected depth in every configuration, and that agreement tightens with prior strength. Both sets include the injection law, so these results show that a model average tracks truth when at least one tested LDL can represent truth.}}
    \label{fig:fig7}
\end{figure*}

Fourth, the fourth-order nonlinear LDL performs best. Functionally matching the injection model, it results in the lowest transit depth bias and amplification factor. Tightening \vtwo{from uniform priors to Gaussian priors with 10--20\% widths} reduces $A$ by a factor of \vtwo{$1.1$--$1.8$}, and pushing priors further to $1\%$ \vtwo{widths}---potentially achievable with future stellar models---yields a \vtwo{$2.1$--$2.5\times$} decrease. Such gains in transit depth precision translate to requiring \vtwo{$\sim1$--$3\times$ and $\sim4$--$6\times$} fewer transits to achieve the same level of scientific return (e.g., placing constraints on atmospheric properties from transmission spectra).

\vtwo{Finally, the trends with prior strength reveal a tension. Tighter priors consistently reduce $A$, as expected, since narrower priors constrain both the LDC and transit depth posteriors, with two exceptions. The first concerns quadratic fits between $5\%$ and $1\%$ prior widths, discussed earlier. The second concerns the third-order polynomial LDL, for which $A$ is roughly constant across uniform, $20\%$, and $10\%$ prior widths but rises at $5\%$ and $1\%$. The mechanism is the same in both cases: as priors tighten, the LDCs can no longer move to the optimum that best recovers $D$, so the orbital parameters must compensate, increasing their correlations with $D$ and broadening its posterior. The transition that is setting in at wider priors for the third-order law suggests its retrievals depend more heavily on LDC--orbital coadjustments, such that constraining the LDCs removes more of the fit's capacity to compensate than for quadratic fits. Why this effect is confined to these two laws is not immediately apparent and warrants further investigation.}

\vtwo{The corresponding impact on bias is less clear. The same mismatch drives it, but unlike $A$, bias does not respond monotonically to prior width and instead depends on the LDL selected. For instance, for the fifth-order polynomial, bias grows steadily as priors tighten, whereas the third-order law improves down to $10\%$ before worsening again, and the sixth-order law reverses twice. Three effects likely contribute. First, prior-induced biases on different LDCs need not point in the same direction, so their net effect on $D$ varies non-trivially between configurations. Second, bias is reported in units of $\sigma_D$, which is itself altered by prior choice: where tightening inflates $\sigma_D$ (e.g., for the third-order law below $10\%$), a growing absolute $D$ bias can appear unchanged, or even reduced, once normalized. Third, with 10 noise realizations per configuration, we have limited power to distinguish genuine reversals from scatter. The fourth-order nonlinear LDL is immune to this tension since it functionally matches the injection model, meaning tighter priors reduce $A$ without inflating bias.}

Taken together, these results suggest that observers should fit modern light curves with both a third-order polynomial and fourth-order nonlinear LDL. Both prescriptions provide an optimal compromise between $A$ and bias, while also serving as useful cross-checks: their distinct functional forms make them sensitive to different aspects of the stellar intensity profile, enabling a more complete characterization of the underlying stellar properties. As a community, we should therefore encourage the use of multiple LDLs when fitting light curves and recommend reporting results from each prescription to highlight whether transit depth measurements differ or remain consistent across models. Particular care should also be paid to the wavelength dependence of limb-darkening \vtwo{parameterization} (exemplified in \autoref{fig:appendix_fig1}), since inadequate choices can introduce spurious features in transmission spectra that mimic genuine atmospheric signals \citep[e.g.,][]{Keers2024}, \vtwo{as we demonstrate in \autoref{sec:chrom_transit_depth_bias}.}

\subsection{\vtwo{Chromatic transit depth bias}}\label{sec:chrom_transit_depth_bias}

\vtwo{The cluster modes of \autoref{sec:impact_prior_choice} differ simultaneously across four parameters, and so their results cannot on their own isolate the sought-after wavelength dependence. In an effort to do just that, and mirroring the approach of \cite{Keers2024}, we perform injection--retrieval tests over chromatic light curves and extract a mock transmission spectrum.}

\vtwo{For five cluster modes spanning a broad range of $(T_{\rm eff}, \log g, [{\rm M/H}])$-parameter space, we use \texttt{ExoTiC-LD} to retrieve stellar intensity profiles at each wavelength of an $R=100$ grid across the NIRSpec/PRISM range, and we fit each profile as in \autoref{sec:appropriate_coeff} to obtain a chromatic set of injection LDCs. The orbital parameters and transit depth are kept achromatic by construction, so that any wavelength dependence in the resulting spectrum originates solely from limb darkening. Each light curve is fit with three laws: the quadratic, third-order polynomial, and fourth-order nonlinear law. We retain \autoref{tab:tab1}'s broad uniform priors and starting positions for the LDCs and transit depth, but impose tight Gaussian priors on the orbital parameters informed by an initial white light curve fit, as is common practice. As the orbital parameters are already well constrained, we reduce the MCMC chain length by an order of magnitude to $10{,}000$ steps, discarding the first $7{,}000$ as burn-in, and post-process as before.}

\vtwo{\autoref{fig:fig6} shows each cluster mode's transmission spectrum. Echoing \cite{Keers2024}, the quadratic law imprints the most structure, with peak-to-peak amplitudes $>40$ ppm in every stellar regime, while the third-order law reduces this to below $30$ ppm, and the injection law returns a flat spectrum, within the noise level. In the context of atmospheric studies, such a wavelength-dependent bias is non-negligible and could be mistaken for atmospheric structure.}

\vtwo{\autoref{fig:appendix_fig7} reports the corresponding amplification factors. $A$ decreases across the PRISM range for all three laws by a factor $1.3$--$3.2$, with most of the decline happening before $1.5\,\mu$m.\footnote{Tight priors are placed on the orbital parameters in these retrievals, so $A$ values are systematically lower than those of \autoref{fig:fig5}, where the orbital parameters are free. The two should be compared in trend rather than in absolute value.} Its chromatic features mirror those of the bias, most notably for the quadratic law: where the bias is largest, so is $A$. Interestingly, below $2\,\mu$m, the quadratic law can produce higher $A$ than the third-order polynomial law, losing even the precision advantage that would otherwise motivate its use. Overall, the third-order polynomial is the best compromise, consistently keeping the chromatic bias below $30$ ppm and the amplification factor $\leq 6$.}

\subsection{\vtwo{Comparison with empirical results}}\label{sec:compare_empirics}
\vtwo{
Our conclusions are in tension with \citetalias{Sing2026}, who found the quadratic law is statistically preferred for six of seven JWST targets and induces negligible transit depth bias of $\sim14$\,ppm. To shed light on this gap, we investigate two possible sources.}

\vtwo{The first is the reference point. Our biases are measured against the injected truth, whereas \citetalias{Sing2026} compared inferred transit depth against the Bayesian model average across four different LDLs. Whether this is a sound proxy for truth is testable in our framework. From the injection--retrievals of \autoref{fig:fig5}, we computed the BMA for every combination of stellar parameters, prior strength, and noise realization, both over all LDLs considered and over a restricted set closest to the LDLs of \citetalias{Sing2026}. As illustrated in \autoref{fig:fig7}, the BMA is within $10$ ppm of the injected depth for every configuration. However, in our tests the candidate set always contains the fourth-order nonlinear law used for injection, so at least one member exactly represents the true intensity profile. Thus, our results do not establish BMA as an unbiased reference for general use; rather it is a useful diagnostic of fit sensitivity to limb-darkening prescription. It also means that the ${\sim}14$\,ppm of \citetalias{Sing2026}, a white-light comparison against their own average, is not a measurement of accuracy against known truth, and so is not directly comparable to the biases we report.}

\vtwo{The second difference concerns how the LDCs are derived. Excluding the intensities closest to the limb is a long-standing practice: \cite{Sing2010} adopted a $\mu > 0.05$ cutoff after finding it improved agreement between model and empirical LDCs for the first seven CoRoT planets, and similar truncations have since been applied by \cite{Claret2018} and \cite{Maxted2023}, motivated by the fact that stellar models over-predict limb darkening. \citetalias{Sing2026} pushed this further, reporting that achromatic offsets between model and empirical LDCs are minimized at $\mu_{\rm min}=0.2$.}

\vtwo{Throughout our analyses, we have fit down to $\mu=0.01$, which begs the question: are our results driven by the models' heightened limb darkening? We test this in \autoref{sec:appendix_mu_min} by, first, regenerating the LDCs for \autoref{sec:impact_prior_choice}'s modes with fits over a 100-point $\mu$ grid spanning $[0.2, 1.0]$, before reproducing \autoref{fig:fig5}'s first panel in \autoref{fig:appendix_fig8}. As illustrated, the conclusions are unaffected: the quadratic law still yields high bias and low $A$, high-order laws give low bias and high $A$, and the third-order polynomial still provides the best compromise.}

\vtwo{\autoref{fig:appendix_fig9} underpins these results: above $\mu=0.2$ the two fits are indistinguishable, diverging only minimally below it. The half-integer terms that the quadratic law cannot represent are therefore not an artifact of the over-darkened limb, but an inherent property of the bulk of the intensity profile. This is consistent with transit depth bias being governed by the LDC--$D$ correlations identified in \autoref{sec:amp_factor_cause}. We stress that this test does not establish that stellar model intensities are accurate near the limb, only that increasing $\mu_{\rm min}$ does not impact the bias incurred by the choice of LDL.}

\subsection{\vtwo{Beyond limb darkening}}\label{sec:orbital_params}
\vtwo{The mitigation strategies we have explored specifically target the LDCs. This reflects the conclusions of \autoref{sec:amp_factor_cause}, where they emerged as the dominant contributors to the TDPP, but it leaves a second lever unexamined. The sensitivity analysis identified significant correlations between the inclination and $D$, while the ablation study found that fixing $i$ to truth changed $A$ by $\sim20\%$---smaller than the $\sim30$--$50\%$ obtained by fixing the LDCs, but far from negligible.}

\vtwo{Extending the prior sweep of \autoref{sec:impact_prior_choice} to $i$ is therefore a natural direction for future work. The key difference lies in where such priors would come from: whereas those on LDCs were drawn from stellar atmosphere models, constraints on $i$ must come from complementary observations. One avenue is Rossiter--McLaughlin observations \citep{Rossiter1924, McLaughlin1924}, whose amplitude scales to first order as $\sqrt{1-b^2}$ \citep{Triaud2018} and would thus constrain the impact parameter independently from transit photometry. Because $b$, $a/R_\star$, and $i$ are strongly covariant in photometric fits, such an external constraint would tighten this part of the posterior and, in turn, the inclination. However, how much of the residual amplification it recovers remains to be quantified.}

\section{Conclusion}\label{sec:conclusion}
\vtwo{We have examined why exoplanet transit depth measurements ($D$) can remain less precise---and sometimes less accurate---than theoretical expectations, a shortfall we define as the Transit Depth Precision Problem (TDPP). We quantify the precision loss with the amplification factor $A = \sigma_D/\sigma_{\rm ref}$ which systematically exceeds its theoretical floor of $\mathbf{\sqrt{3/2}}$. In our injection--retrievals, $A$ is set by degeneracies among $D$, the LDCs, and the orbital parameters: fixing individual LDCs to truth produces the largest changes in $A$ ($\sim30$--$50\%$), followed by the inclination and stellar density ($\sim20$--$25\%$). These sensitivities are not additive, but together they establish that the excess uncertainty is not a purely instrumental limitation.}

\vtwo{Precision and accuracy must be considered together: choices that reduce $A$, such as tightening LDC priors or lowering the LDL order, inflate the bias on $D$, and vice versa. The quadratic law exemplifies this, achieving $A\sim2$--$6$ while producing biases of $2\sigma$--$30\sigma$, and wavelength-resolved experiments show it imprints spurious chromatic structure on an intrinsically flat transmission spectrum with a peak-to-peak amplitude $>40$ ppm. We instead favor the third-order polynomial and fourth-order nonlinear laws, which at the $10\%$--$20\%$ prior widths currently achievable from stellar models, yield $A \sim 3$--$6$ and biases of $2\sigma$--$3\sigma$, falling below $2\sigma$ for $10\%$. Because the severity of the quadratic law's inadequacy varies markedly across the ($T_{\rm eff}$, $\log g$, $[{\rm M/H}]$, $\lambda$)-space probed, its suitability should be verified per target and bandpass rather than assumed. These conclusions remain conditional on the MPS-ATLAS intensity profiles and central-transit geometry explored here.}

\vtwo{Since bias is inaccessible in real analyses, we recommend two diagnostics.} First, report $A$ in each wavelength bin: $A\gg6$ suggests tighter constraints are accessible, while $A\ll6$ implies artificially tight constraints and potentially biased measurements. \vtwo{Second, fit transits with several, flexible LDLs and report both the individual transit depths and a model average. In our tests, the Bayesian model-averaged depth lies within 10 ppm of truth whenever the candidate set of laws is expressive enough to represent the true stellar intensity profile. The spread among candidate models measures sensitivity to limb-darkening prescription, though it cannot reveal a deficiency shared by every law tested.}

\vtwo{The largest practical gain would come from improving the stellar models that inform LDC priors. In our fourth-order nonlinear tests, tightening priors from uniform to $10\%$--$20\%$ widths reduce $\sigma_D$ by factors of $1.1$--$1.8$ at fixed data quality, and $1\%$ widths by $2.1$--$2.5$. Assuming $\sigma_D$ decreases as the inverse square root of the number of observed transits, these gains correspond to reaching the same target precision with ${\sim}1$--$3$ and ${\sim}4$--$6$ times fewer transits, respectively. These are conditional projections: realizing them requires priors that are not only narrower but accurately centered, since a systematic bias does not average away by observing more transits.}

\vtwo{Longer term, higher-fidelity stellar atmospheric models, limb-darkening parameterizations designed to reduce $D$--LDC degeneracies \citep[e.g., more orthogonal bases, as in][]{Maxted2018,Maxted2023}, and external constraints on the correlated orbital parameters---for instance Rossiter--McLaughlin observations, which constrain the impact parameter independently of transit photometry---offer a path toward improving accuracy and precision together and, therefore, extracting more information from each transit.}

\section*{Acknowledgments}

We thank the reviewer for a careful and constructive review, which has substantially strengthened the manuscript.
This work was funded by the European Research Council (ERC) under the European Union’s Horizon 2020 research and innovation program (grant No.\ 101118581).
This material is based upon work supported by the National Aeronautics and Space Administration under Agreement No.\ 80NSSC21K0593 for the program ``Alien Earths''.
The results reported herein benefited from collaborations and/or information exchange within NASA’s Nexus for Exoplanet System Science (NExSS) research coordination network sponsored by NASA’s Science Mission Directorate.
SJM acknowledges support from the Massachusetts Institute of Technology through the Praecis Presidential Fellowship, Desmond Fellowship, and Hurley Fellowship.
\software{
\texttt{arviz}\footnote{\url{https://www.arviz.org/en/latest/}}\protect\citep[v0.23.4,][]{arviz},
\texttt{astropy}\footnote{\url{https://www.astropy.org}}\protect\citep[v7.2.0,][]{astropy_2013,astropy_2018,astropy_2022},
\texttt{corner}\footnote{\url{https://corner.readthedocs.io/en/latest/}}\protect\citep[v2.2.2,][]{corner},
\texttt{emcee-jax}\footnote{\url{https://github.com/SamMerc/emcee-jax}} (v0.1.dev38),
\texttt{ExoTiC-LD}\footnote{\url{https://exotic-ld.readthedocs.io/en/latest/}}\protect\citep[v3.2.0,][]{ExoTiC-LD},
\texttt{jax}\footnote{\url{https://docs.jax.dev/en/latest/}}\protect\citep[v0.10.2,][]{jax},
\texttt{jaxoplanet}\footnote{\url{https://jax.exoplanet.codes/en/latest/}}\protect\citep[v0.1.0,][]{jaxoplanet},
\texttt{lmfit}\footnote{\url{https://lmfit.github.io/lmfit-py/}}\protect\citep[v1.3.3,][]{lmfit},
\texttt{Matplotlib}\footnote{\url{https://matplotlib.org}}\protect\citep[v3.10.9,][]{Matplotlib},
\texttt{numpy}\footnote{\url{https://numpy.org}}\protect\citep[v2.4.6,][]{numpy},
\texttt{numpyro}\footnote{\url{https://num.pyro.ai/en/stable/}}\protect\citep[v0.19.0,][]{numpyro},
\texttt{pandas}\footnote{\url{https://pandas.pydata.org}}\protect\citep[v2.2.3,][]{pandas},
\texttt{scikit-learn}\footnote{\url{https://scikit-learn.org/stable/}}\protect\citep[v1.6.1,][]{scikit-learn},
\texttt{scipy}\footnote{\url{https://scipy.org}}\protect\citep[v1.17.1,][]{scipy},
\texttt{squishyplanet}\footnote{\url{https://squishyplanet.readthedocs.io/en/latest/}}\protect\citep[v0.3.1,][]{squishyplanet},
\texttt{tqdm}\footnote{\url{https://tqdm.github.io}}\protect\citep[v4.67.1,][]{tqdm}
}

\bibliography{bib}

\begin{thebibliography}{}
\expandafter\ifx\csname natexlab\endcsname\relax\def\natexlab#1{#1}\fi
\providecommand{\url}[1]{\href{#1}{#1}}
\providecommand{\dodoi}[1]{doi:~\href{http://doi.org/#1}{\nolinkurl{#1}}}
\providecommand{\doeprint}[1]{\href{http://ascl.net/#1}{\nolinkurl{http://ascl.net/#1}}}
\providecommand{\doarXiv}[1]{\href{https://arxiv.org/abs/#1}{\nolinkurl{https://arxiv.org/abs/#1}}}

\bibitem[{{Astropy Collaboration} {et~al.}(2013){Astropy Collaboration},
  {Robitaille}, {Tollerud}, {Greenfield}, {Droettboom}, {Bray}, {Aldcroft},
  {Davis}, {Ginsburg}, {Price-Whelan}, {Kerzendorf}, {Conley}, {Crighton},
  {Barbary}, {Muna}, {Ferguson}, {Grollier}, {Parikh}, {Nair}, {Unther},
  {Deil}, {Woillez}, {Conseil}, {Kramer}, {Turner}, {Singer}, {Fox}, {Weaver},
  {Zabalza}, {Edwards}, {Azalee Bostroem}, {Burke}, {Casey}, {Crawford},
  {Dencheva}, {Ely}, {Jenness}, {Labrie}, {Lim}, {Pierfederici}, {Pontzen},
  {Ptak}, {Refsdal}, {Servillat}, \& {Streicher}}]{astropy_2013}
{Astropy Collaboration}, {Robitaille}, T.~P., {Tollerud}, E.~J., {et~al.} 2013,
  \aap, 558, A33, \dodoi{10.1051/0004-6361/201322068}

\bibitem[{{Astropy Collaboration} {et~al.}(2018){Astropy Collaboration},
  {Price-Whelan}, {Sip{\H{o}}cz}, {G{\"u}nther}, {Lim}, {Crawford}, {Conseil},
  {Shupe}, {Craig}, {Dencheva}, {Ginsburg}, {Vand erPlas}, {Bradley},
  {P{\'e}rez-Su{\'a}rez}, {de Val-Borro}, {Aldcroft}, {Cruz}, {Robitaille},
  {Tollerud}, {Ardelean}, {Babej}, {Bach}, {Bachetti}, {Bakanov}, {Bamford},
  {Barentsen}, {Barmby}, {Baumbach}, {Berry}, {Biscani}, {Boquien}, {Bostroem},
  {Bouma}, {Brammer}, {Bray}, {Breytenbach}, {Buddelmeijer}, {Burke},
  {Calderone}, {Cano Rodr{\'\i}guez}, {Cara}, {Cardoso}, {Cheedella}, {Copin},
  {Corrales}, {Crichton}, {D'Avella}, {Deil}, {Depagne}, {Dietrich}, {Donath},
  {Droettboom}, {Earl}, {Erben}, {Fabbro}, {Ferreira}, {Finethy}, {Fox},
  {Garrison}, {Gibbons}, {Goldstein}, {Gommers}, {Greco}, {Greenfield},
  {Groener}, {Grollier}, {Hagen}, {Hirst}, {Homeier}, {Horton}, {Hosseinzadeh},
  {Hu}, {Hunkeler}, {Ivezi{\'c}}, {Jain}, {Jenness}, {Kanarek}, {Kendrew},
  {Kern}, {Kerzendorf}, {Khvalko}, {King}, {Kirkby}, {Kulkarni}, {Kumar},
  {Lee}, {Lenz}, {Littlefair}, {Ma}, {Macleod}, {Mastropietro}, {McCully},
  {Montagnac}, {Morris}, {Mueller}, {Mumford}, {Muna}, {Murphy}, {Nelson},
  {Nguyen}, {Ninan}, {N{\"o}the}, {Ogaz}, {Oh}, {Parejko}, {Parley}, {Pascual},
  {Patil}, {Patil}, {Plunkett}, {Prochaska}, {Rastogi}, {Reddy Janga},
  {Sabater}, {Sakurikar}, {Seifert}, {Sherbert}, {Sherwood-Taylor}, {Shih},
  {Sick}, {Silbiger}, {Singanamalla}, {Singer}, {Sladen}, {Sooley},
  {Sornarajah}, {Streicher}, {Teuben}, {Thomas}, {Tremblay}, {Turner},
  {Terr{\'o}n}, {van Kerkwijk}, {de la Vega}, {Watkins}, {Weaver}, {Whitmore},
  {Woillez}, {Zabalza}, \& {Astropy Contributors}}]{astropy_2018}
{Astropy Collaboration}, {Price-Whelan}, A.~M., {Sip{\H{o}}cz}, B.~M., {et~al.}
  2018, \aj, 156, 123, \dodoi{10.3847/1538-3881/aabc4f}

\bibitem[{{Astropy Collaboration} {et~al.}(2022){Astropy Collaboration},
  {Price-Whelan}, {Lim}, {Earl}, {Starkman}, {Bradley}, {Shupe}, {Patil},
  {Corrales}, {Brasseur}, {N{"o}the}, {Donath}, {Tollerud}, {Morris},
  {Ginsburg}, {Vaher}, {Weaver}, {Tocknell}, {Jamieson}, {van Kerkwijk},
  {Robitaille}, {Merry}, {Bachetti}, {G{"u}nther}, {Aldcroft},
  {Alvarado-Montes}, {Archibald}, {B{'o}di}, {Bapat}, {Barentsen}, {Baz{'a}n},
  {Biswas}, {Boquien}, {Burke}, {Cara}, {Cara}, {Conroy}, {Conseil}, {Craig},
  {Cross}, {Cruz}, {D'Eugenio}, {Dencheva}, {Devillepoix}, {Dietrich},
  {Eigenbrot}, {Erben}, {Ferreira}, {Foreman-Mackey}, {Fox}, {Freij}, {Garg},
  {Geda}, {Glattly}, {Gondhalekar}, {Gordon}, {Grant}, {Greenfield}, {Groener},
  {Guest}, {Gurovich}, {Handberg}, {Hart}, {Hatfield-Dodds}, {Homeier},
  {Hosseinzadeh}, {Jenness}, {Jones}, {Joseph}, {Kalmbach}, {Karamehmetoglu},
  {Ka{l}uszy{'n}ski}, {Kelley}, {Kern}, {Kerzendorf}, {Koch}, {Kulumani},
  {Lee}, {Ly}, {Ma}, {MacBride}, {Maljaars}, {Muna}, {Murphy}, {Norman},
  {O'Steen}, {Oman}, {Pacifici}, {Pascual}, {Pascual-Granado}, {Patil},
  {Perren}, {Pickering}, {Rastogi}, {Roulston}, {Ryan}, {Rykoff}, {Sabater},
  {Sakurikar}, {Salgado}, {Sanghi}, {Saunders}, {Savchenko}, {Schwardt},
  {Seifert-Eckert}, {Shih}, {Jain}, {Shukla}, {Sick}, {Simpson},
  {Singanamalla}, {Singer}, {Singhal}, {Sinha}, {Sip{H{o}}cz}, {Spitler},
  {Stansby}, {Streicher}, {{{S}}umak}, {Swinbank}, {Taranu}, {Tewary},
  {Tremblay}, {Val-Borro}, {Van Kooten}, {Vasovi{'c}}, {Verma}, {de Miranda
  Cardoso}, {Williams}, {Wilson}, {Winkel}, {Wood-Vasey}, {Xue}, {Yoachim},
  {Zhang}, {Zonca}, \& {Astropy Project Contributors}}]{astropy_2022}
{Astropy Collaboration}, {Price-Whelan}, A.~M., {Lim}, P.~L., {et~al.} 2022,
  \apj, 935, 167, \dodoi{10.3847/1538-4357/ac7c74}

\bibitem[{{Benneke} \& {Seager}(2012)}]{Benneke2012}
{Benneke}, B., \& {Seager}, S. 2012, \apj, 753, 100,
  \dodoi{10.1088/0004-637X/753/2/100}

\bibitem[{{Benneke} {et~al.}(2024){Benneke}, {Roy}, {Coulombe}, {Radica},
  {Piaulet}, {Ahrer}, {Pierrehumbert}, {Krissansen-Totton}, {Schlichting},
  {Hu}, {Yang}, {Christie}, {Thorngren}, {Young}, {Pelletier}, {Knutson},
  {Miguel}, {Evans-Soma}, {Dorn}, {Gagnebin}, {Fortney}, {Komacek},
  {MacDonald}, {Raul}, {Cloutier}, {Acuna}, {Lafreni{\`e}re}, {Cadieux},
  {Doyon}, {Welbanks}, \& {Allart}}]{Benneke2024}
{Benneke}, B., {Roy}, P.-A., {Coulombe}, L.-P., {et~al.} 2024, arXiv e-prints,
  arXiv:2403.03325, \dodoi{10.48550/arXiv.2403.03325}

\bibitem[{{Berardo} \& {de Wit}(2022)}]{Berardo2022}
{Berardo}, D., \& {de Wit}, J. 2022, \apj, 935, 178,
  \dodoi{10.3847/1538-4357/ac82b2}

\bibitem[{{Bonfils} {et~al.}(2013){Bonfils}, {Delfosse}, {Udry}, {Forveille},
  {Mayor}, {Perrier}, {Bouchy}, {Gillon}, {Lovis}, {Pepe}, \&
  et~al.}]{Bonfils2013}
{Bonfils}, X., {Delfosse}, X., {Udry}, S., {et~al.} 2013, \aap, 549, A109,
  \dodoi{10.1051/0004-6361/201014704}

\bibitem[{{Bourque} {et~al.}(2021){Bourque}, {Espinoza}, {Filippazzo}, {Fix},
  {King}, {Martlin}, {Medina}, {Batalha}, {Fox}, {Fowler}, {Fraine}, {Hill},
  {Lewis}, {Stevenson}, {Valenti}, \& {Wakeford}}]{ExoCTK}
{Bourque}, M., {Espinoza}, N., {Filippazzo}, J., {et~al.} 2021, {The Exoplanet
  Characterization Toolkit (ExoCTK)}, 1.0.0,  Zenodo,
  \dodoi{10.5281/zenodo.4556063}

\bibitem[{Bradbury {et~al.}(2018)Bradbury, Frostig, Hawkins, Johnson, Katariya,
  Leary, Maclaurin, Necula, Paszke, Vander{P}las, Wanderman-{M}ilne, \&
  Zhang}]{jax}
Bradbury, J., Frostig, R., Hawkins, P., {et~al.} 2018, {JAX}: composable
  transformations of {P}ython+{N}um{P}y programs, 0.3.13.
\newblock \url{http://github.com/jax-ml/jax}

\bibitem[{{Carter} {et~al.}(2011){Carter}, {Winn}, {Holman}, {Fabrycky},
  {Berta}, {Burke}, \& {Nutzman}}]{Carter2011}
{Carter}, J.~A., {Winn}, J.~N., {Holman}, M.~J., {et~al.} 2011, \apj, 730, 82,
  \dodoi{10.1088/0004-637X/730/2/82}

\bibitem[{{Carter} {et~al.}(2008){Carter}, {Yee}, {Eastman}, {Gaudi}, \&
  {Winn}}]{Carter2008}
{Carter}, J.~A., {Yee}, J.~C., {Eastman}, J., {Gaudi}, B.~S., \& {Winn}, J.~N.
  2008, \apj, 689, 499, \dodoi{10.1086/592321}

\bibitem[{{Cassese} {et~al.}(2024){Cassese}, {Vega}, {Lu}, {Rice}, {Poddar}, \&
  {Kipping}}]{squishyplanet}
{Cassese}, B., {Vega}, J., {Lu}, T., {et~al.} 2024, The Journal of Open Source
  Software, 9, 6972, \dodoi{10.21105/joss.06972}

\bibitem[{{Charbonneau} {et~al.}(2000){Charbonneau}, {Brown}, {Latham}, \&
  {Mayor}}]{Charbonneau2000}
{Charbonneau}, D., {Brown}, T.~M., {Latham}, D.~W., \& {Mayor}, M. 2000, \apjl,
  529, L45, \dodoi{10.1086/312457}

\bibitem[{{Claret}(1998)}]{Claret1998}
{Claret}, A. 1998, \aap, 335, 647

\bibitem[{{Claret}(2000)}]{Claret2000}
---. 2000, \aap, 363, 1081

\bibitem[{{Claret}(2018)}]{Claret2018}
---. 2018, \aap, 618, A20, \dodoi{10.1051/0004-6361/201833060}

\bibitem[{{Claret} \& {Bloemen}(2011)}]{Claret2011}
{Claret}, A., \& {Bloemen}, S. 2011, \aap, 529, A75,
  \dodoi{10.1051/0004-6361/201116451}

\bibitem[{{Csizmadia} {et~al.}(2013){Csizmadia}, {Pasternacki}, {Dreyer},
  {Cabrera}, {Erikson}, \& {Rauer}}]{Csizmadia2013}
{Csizmadia}, S., {Pasternacki}, T., {Dreyer}, C., {et~al.} 2013, \aap, 549, A9,
  \dodoi{10.1051/0004-6361/201219888}

\bibitem[{{da Costa-Luis}(2019)}]{tqdm}
{da Costa-Luis}, C. 2019, The Journal of Open Source Software, 4, 1277,
  \dodoi{10.21105/joss.01277}

\bibitem[{{de Wit} {et~al.}(2012){de Wit}, {Gillon}, {Demory}, \&
  {Seager}}]{deWit2012}
{de Wit}, J., {Gillon}, M., {Demory}, B.~O., \& {Seager}, S. 2012, \aap, 548,
  A128, \dodoi{10.1051/0004-6361/201219060}

\bibitem[{{de Wit} \& {Seager}(2013)}]{deWit2013}
{de Wit}, J., \& {Seager}, S. 2013, Science, 342, 1473,
  \dodoi{10.1126/science.1245450}

\bibitem[{{Deming} {et~al.}(2009){Deming}, {Seager}, {Winn}, {Miller-Ricci},
  {Clampin}, {Lindler}, {Greene}, {Charbonneau}, {Laughlin}, {Ricker},
  {Latham}, \& {Ennico}}]{Deming2009}
{Deming}, D., {Seager}, S., {Winn}, J., {et~al.} 2009, \pasp, 121, 952,
  \dodoi{10.1086/605913}

\bibitem[{{Dumusque} {et~al.}(2011{\natexlab{a}}){Dumusque}, {Santos}, {Udry},
  {Lovis}, \& {Bonfils}}]{Dumusque2011b}
{Dumusque}, X., {Santos}, N.~C., {Udry}, S., {Lovis}, C., \& {Bonfils}, X.
  2011{\natexlab{a}}, \aap, 527, A82, \dodoi{10.1051/0004-6361/201015877}

\bibitem[{{Dumusque} {et~al.}(2011{\natexlab{b}}){Dumusque}, {Udry}, {Lovis},
  {Santos}, \& {Monteiro}}]{Dumusque2011a}
{Dumusque}, X., {Udry}, S., {Lovis}, C., {Santos}, N.~C., \& {Monteiro},
  M.~J.~P.~F.~G. 2011{\natexlab{b}}, \aap, 525, A140,
  \dodoi{10.1051/0004-6361/201014097}

\bibitem[{{Dyrek} {et~al.}(2024){Dyrek}, {Min}, {Decin}, {Bouwman}, {Crouzet},
  {Molli{\`e}re}, {Lagage}, {Konings}, {Tremblin}, {G{\"u}del}, {Pye},
  {Waters}, {Henning}, {Vandenbussche}, {Ardevol Martinez}, {Argyriou},
  {Ducrot}, {Heinke}, {van Looveren}, {Absil}, {Barrado}, {Baudoz},
  {Boccaletti}, {Cossou}, {Coulais}, {Edwards}, {Gastaud}, {Glasse}, {Glauser},
  {Greene}, {Kendrew}, {Krause}, {Lahuis}, {Mueller}, {Olofsson}, {Patapis},
  {Rouan}, {Royer}, {Scheithauer}, {Waldmann}, {Whiteford}, {Colina}, {van
  Dishoeck}, {{\"O}stlin}, {Ray}, \& {Wright}}]{Dyrek2024}
{Dyrek}, A., {Min}, M., {Decin}, L., {et~al.} 2024, \nat, 625, 51,
  \dodoi{10.1038/s41586-023-06849-0}

\bibitem[{{Espinoza} \& {Jord{\'a}n}(2015)}]{Espinoza2015}
{Espinoza}, N., \& {Jord{\'a}n}, A. 2015, \mnras, 450, 1879,
  \dodoi{10.1093/mnras/stv744}

\bibitem[{{Espinoza} \& {Jord{\'a}n}(2016)}]{Espinoza2016}
---. 2016, \mnras, 457, 3573, \dodoi{10.1093/mnras/stw224}

\bibitem[{Everitt {et~al.}(2011)Everitt, Landau, Leese, \& Stahl}]{Everitt2011}
Everitt, B.~S., Landau, S., Leese, M., \& Stahl, D. 2011, Cluster Analysis, 5th
  edn. (Wiley), \dodoi{10.1002/9780470977811}

\bibitem[{Foreman-Mackey(2016)}]{corner}
Foreman-Mackey, D. 2016, The Journal of Open Source Software, 1, 24,
  \dodoi{10.21105/joss.00024}

\bibitem[{{Foreman-Mackey} {et~al.}(2013){Foreman-Mackey}, {Hogg}, {Lang}, \&
  {Goodman}}]{emcee}
{Foreman-Mackey}, D., {Hogg}, D.~W., {Lang}, D., \& {Goodman}, J. 2013, \pasp,
  125, 306, \dodoi{10.1086/670067}

\bibitem[{{Gim{\'e}nez}(2006)}]{Gimenez2006}
{Gim{\'e}nez}, A. 2006, \aap, 450, 1231, \dodoi{10.1051/0004-6361:20054445}

\bibitem[{Grant \& Wakeford(2024)}]{ExoTiC-LD}
Grant, D., \& Wakeford, H.~R. 2024, Journal of Open Source Software, 9, 6816,
  \dodoi{10.21105/joss.06816}

\bibitem[{{Greene} {et~al.}(2016){Greene}, {Line}, {Montero}, {Fortney},
  {Lustig-Yaeger}, \& {Luther}}]{Greene2016}
{Greene}, T.~P., {Line}, M.~R., {Montero}, C., {et~al.} 2016, \apj, 817, 17,
  \dodoi{10.3847/0004-637X/817/1/17}

\bibitem[{Harris {et~al.}(2020)Harris, Millman, van~der Walt, Gommers,
  Virtanen, Cournapeau, Wieser, Taylor, Berg, Smith, Kern, Picus, Hoyer, van
  Kerkwijk, Brett, Haldane, del R{\'{i}}o, Wiebe, Peterson,
  G{\'{e}}rard-Marchant, Sheppard, Reddy, Weckesser, Abbasi, Gohlke, \&
  Oliphant}]{numpy}
Harris, C.~R., Millman, K.~J., van~der Walt, S.~J., {et~al.} 2020, Nature, 585,
  357, \dodoi{10.1038/s41586-020-2649-2}

\bibitem[{Hattori {et~al.}(2024)Hattori, Garcia, Murray, Dong, Dholakia, Degen,
  \& Foreman-Mackey}]{jaxoplanet}
Hattori, S., Garcia, L., Murray, C., {et~al.} 2024, {exoplanet-dev/jaxoplanet:
  Astronomical time series analysis with JAX}, v0.0.2,  Zenodo,
  \dodoi{10.5281/zenodo.10736936}

\bibitem[{{Hayek} {et~al.}(2012){Hayek}, {Sing}, {Pont}, \&
  {Asplund}}]{Hayek2012}
{Hayek}, W., {Sing}, D., {Pont}, F., \& {Asplund}, M. 2012, \aap, 539, A102,
  \dodoi{10.1051/0004-6361/201117868}

\bibitem[{{Henry} {et~al.}(2000){Henry}, {Marcy}, {Butler}, \&
  {Vogt}}]{Henry2000}
{Henry}, G.~W., {Marcy}, G.~W., {Butler}, R.~P., \& {Vogt}, S.~S. 2000, \apjl,
  529, L41, \dodoi{10.1086/312458}

\bibitem[{{Howarth}(2011)}]{Howarth2011}
{Howarth}, I.~D. 2011, \mnras, 418, 1165,
  \dodoi{10.1111/j.1365-2966.2011.19568.x}

\bibitem[{Hunter(2007)}]{Matplotlib}
Hunter, J.~D. 2007, Computing in Science \& Engineering, 9, 90,
  \dodoi{10.1109/MCSE.2007.55}

\bibitem[{{Husser} {et~al.}(2013){Husser}, {Wende-von Berg}, {Dreizler},
  {Homeier}, {Reiners}, {Barman}, \& {Hauschildt}}]{Husser2013}
{Husser}, T.~O., {Wende-von Berg}, S., {Dreizler}, S., {et~al.} 2013, \aap,
  553, A6, \dodoi{10.1051/0004-6361/201219058}

\bibitem[{{Keers} {et~al.}(2024){Keers}, {Shapiro}, {Kostogryz}, {Glidden},
  {Niraula}, {Rackham}, {Seager}, {Solanki}, {Unruh}, {Vasilyev}, \& {de
  Wit}}]{Keers2024}
{Keers}, R.~E., {Shapiro}, A.~I., {Kostogryz}, N.~M., {et~al.} 2024, \apjl,
  977, L7, \dodoi{10.3847/2041-8213/ad8b51}

\bibitem[{{Kipping}(2013)}]{Kipping2013}
{Kipping}, D.~M. 2013, \mnras, 435, 2152, \dodoi{10.1093/mnras/stt1435}

\bibitem[{{Kipping}(2016)}]{Kipping2016}
---. 2016, \mnras, 455, 1680, \dodoi{10.1093/mnras/stv2379}

\bibitem[{Kostogryz {et~al.}(2023)Kostogryz, Shapiro, Witzke, Grant, Wakeford,
  Stevenson, Solanki, \& Gizon}]{mpsset2}
Kostogryz, N., Shapiro, A., Witzke, V., {et~al.} 2023, Research Notes of the
  AAS, 7, 39

\bibitem[{Kostogryz {et~al.}(2022)Kostogryz, Witzke, Shapiro, Solanki, Maxted,
  Kurucz, \& Gizon}]{mpsset1}
Kostogryz, N., Witzke, V., Shapiro, A., {et~al.} 2022, Astronomy \&
  Astrophysics, 666, A60

\bibitem[{{Kurucz}(1979)}]{Kurucz1979}
{Kurucz}, R.~L. 1979, \apjs, 40, 1, \dodoi{10.1086/190589}

\bibitem[{{Liu} {et~al.}(2025){Liu}, {Zhu}, {Zhou}, {Hu}, {Lin}, {Dai},
  {Masuda}, \& {Wang}}]{Liu2025}
{Liu}, Q., {Zhu}, W., {Zhou}, Y., {et~al.} 2025, \aj, 169, 79,
  \dodoi{10.3847/1538-3881/ad9b8c}

\bibitem[{{Magic} {et~al.}(2015){Magic}, {Chiavassa}, {Collet}, \&
  {Asplund}}]{Magic2015}
{Magic}, Z., {Chiavassa}, A., {Collet}, R., \& {Asplund}, M. 2015, \aap, 573,
  A90, \dodoi{10.1051/0004-6361/201423804}

\bibitem[{{Magic} {et~al.}(2013){Magic}, {Collet}, {Asplund}, {Trampedach},
  {Hayek}, {Chiavassa}, {Stein}, \& {Nordlund}}]{Magic2013}
{Magic}, Z., {Collet}, R., {Asplund}, M., {et~al.} 2013, \aap, 557, A26,
  \dodoi{10.1051/0004-6361/201321274}

\bibitem[{{Manduca} {et~al.}(1977){Manduca}, {Bell}, \&
  {Gustafsson}}]{Manduca1977}
{Manduca}, A., {Bell}, R.~A., \& {Gustafsson}, B. 1977, \aap, 61, 809

\bibitem[{Martin {et~al.}(2026)Martin, Abril-Pla, Deklerk, Axen, Carroll,
  Hartikainen, \& Vehtari}]{arviz}
Martin, O.~A., Abril-Pla, O., Deklerk, J., {et~al.} 2026, Journal of Open
  Source Software, 11, 9889, \dodoi{10.21105/joss.09889}

\bibitem[{{Maxted}(2018)}]{Maxted2018}
{Maxted}, P.~F.~L. 2018, \aap, 616, A39, \dodoi{10.1051/0004-6361/201832944}

\bibitem[{{Maxted}(2023)}]{Maxted2023}
{Maxted}, P. F.~L. 2023, \mnras, 519, 3723, \dodoi{10.1093/mnras/stac3741}

\bibitem[{{McLaughlin}(1924)}]{McLaughlin1924}
{McLaughlin}, D.~B. 1924, \apj, 60, 22, \dodoi{10.1086/142826}

\bibitem[{{Meunier} {et~al.}(2017){Meunier}, {Mignon}, \&
  {Lagrange}}]{Meunier2017}
{Meunier}, N., {Mignon}, L., \& {Lagrange}, A.~M. 2017, \aap, 607, A124,
  \dodoi{10.1051/0004-6361/201731017}

\bibitem[{{Milne}(1921)}]{Milne1921}
{Milne}, E.~A. 1921, \mnras, 81, 361, \dodoi{10.1093/mnras/81.5.361}

\bibitem[{{Morello} {et~al.}(2017){Morello}, {Tsiaras}, {Howarth}, \&
  {Homeier}}]{Morello2017}
{Morello}, G., {Tsiaras}, A., {Howarth}, I.~D., \& {Homeier}, D. 2017, \aj,
  154, 111, \dodoi{10.3847/1538-3881/aa8405}

\bibitem[{{Neilson} \& {Lester}(2013{\natexlab{a}})}]{Neilson2013a}
{Neilson}, H.~R., \& {Lester}, J.~B. 2013{\natexlab{a}}, \aap, 554, A98,
  \dodoi{10.1051/0004-6361/201321502}

\bibitem[{{Neilson} \& {Lester}(2013{\natexlab{b}})}]{Neilson2013b}
---. 2013{\natexlab{b}}, \aap, 556, A86, \dodoi{10.1051/0004-6361/201321888}

\bibitem[{{Newville} {et~al.}(2016){Newville}, {Stensitzki}, {Allen}, {Rawlik},
  {Ingargiola}, \& {Nelson}}]{lmfit}
{Newville}, M., {Stensitzki}, T., {Allen}, D.~B., {et~al.} 2016, {Lmfit:
  Non-Linear Least-Square Minimization and Curve-Fitting for Python},
  Astrophysics Source Code Library, record ascl:1606.014.
\newblock \doeprint{1606.014}

\bibitem[{{Niraula} {et~al.}(2023){Niraula}, {de Wit}, {Gordon}, {Hargreaves},
  \& {Sousa-Silva}}]{Niraula2023}
{Niraula}, P., {de Wit}, J., {Gordon}, I.~E., {Hargreaves}, R.~J., \&
  {Sousa-Silva}, C. 2023, \apjl, 950, L17, \dodoi{10.3847/2041-8213/acd6f8}

\bibitem[{{Niraula} {et~al.}(2022){Niraula}, {de Wit}, {Gordon}, {Hargreaves},
  {Sousa-Silva}, \& {Kochanov}}]{Niraula2022}
{Niraula}, P., {de Wit}, J., {Gordon}, I.~E., {et~al.} 2022, Nature Astronomy,
  6, 1287, \dodoi{10.1038/s41550-022-01773-1}

\bibitem[{Pedregosa {et~al.}(2011)Pedregosa, Varoquaux, Gramfort, Michel,
  Thirion, Grisel, Blondel, Prettenhofer, Weiss, Dubourg, Vanderplas, Passos,
  Cournapeau, Brucher, Perrot, \& Duchesnay}]{scikit-learn}
Pedregosa, F., Varoquaux, G., Gramfort, A., {et~al.} 2011, Journal of Machine
  Learning Research, 12, 2825

\bibitem[{{Phan} {et~al.}(2019){Phan}, {Pradhan}, \& {Jankowiak}}]{numpyro}
{Phan}, D., {Pradhan}, N., \& {Jankowiak}, M. 2019, arXiv e-prints,
  arXiv:1912.11554, \dodoi{10.48550/arXiv.1912.11554}

\bibitem[{{Rabus} {et~al.}(2009){Rabus}, {Alonso}, {Belmonte}, {Deeg},
  {Gilliland}, {Almenara}, {Brown}, {Charbonneau}, \& {Mandushev}}]{Rabus2009}
{Rabus}, M., {Alonso}, R., {Belmonte}, J.~A., {et~al.} 2009, \aap, 494, 391,
  \dodoi{10.1051/0004-6361:200811110}

\bibitem[{{Rackham} {et~al.}(2017){Rackham}, {Espinoza}, {Apai},
  {L{\'o}pez-Morales}, {Jord{\'a}n}, {Osip}, {Lewis}, {Rodler}, {Fraine},
  {Morley}, \& {Fortney}}]{Rackham2017}
{Rackham}, B., {Espinoza}, N., {Apai}, D., {et~al.} 2017, \apj, 834, 151,
  \dodoi{10.3847/1538-4357/aa4f6c}

\bibitem[{{Rackham} {et~al.}(2018){Rackham}, {Apai}, \&
  {Giampapa}}]{Rackham2018}
{Rackham}, B.~V., {Apai}, D., \& {Giampapa}, M.~S. 2018, \apj, 853, 122,
  \dodoi{10.3847/1538-4357/aaa08c}

\bibitem[{{Rackham} {et~al.}(2019){Rackham}, {Apai}, \&
  {Giampapa}}]{Rackham2019}
---. 2019, \aj, 157, 96, \dodoi{10.3847/1538-3881/aaf892}

\bibitem[{{Rackham} {et~al.}(2023){Rackham}, {Espinoza}, {Berdyugina},
  {Korhonen}, {MacDonald}, {Montet}, {Morris}, {Oshagh}, {Shapiro}, {Unruh},
  {Quintana}, {Zellem}, {Apai}, {Barclay}, {Barstow}, {Bruno}, {Carone},
  {Casewell}, {Cegla}, {Criscuoli}, {Fischer}, {Fournier}, {Giampapa}, {Giles},
  {Iyer}, {Kopp}, {Kostogryz}, {Krivova}, {Mallonn}, {McGruder},
  {Molaverdikhani}, {Newton}, {Panja}, {Peacock}, {Reardon}, {Roettenbacher},
  {Scandariato}, {Solanki}, {Stassun}, {Steiner}, {Stevenson}, {Tregloan-Reed},
  {Valio}, {Wedemeyer}, {Welbanks}, {Yu}, {Alam}, {Davenport}, {Deming},
  {Dong}, {Ducrot}, {Fisher}, {Gilbert}, {Kostov}, {L{\'o}pez-Morales}, {Line},
  {Mo{\v{c}}nik}, {Mullally}, {Paudel}, {Ribas}, \& {Valenti}}]{Rackham2023}
{Rackham}, B.~V., {Espinoza}, N., {Berdyugina}, S.~V., {et~al.} 2023, RAS
  Techniques and Instruments, 2, 148, \dodoi{10.1093/rasti/rzad009}

\bibitem[{{Rossiter}(1924)}]{Rossiter1924}
{Rossiter}, R.~A. 1924, \apj, 60, 15, \dodoi{10.1086/142825}

\bibitem[{{Saba} {et~al.}(2025){Saba}, {Thompson}, {Yip}, {Ma}, {Tsiaras},
  {Al-Refaie}, \& {Tinetti}}]{Saba2025}
{Saba}, A., {Thompson}, A., {Yip}, K.~H., {et~al.} 2025, \apjs, 276, 70,
  \dodoi{10.3847/1538-4365/ad8c3c}

\bibitem[{{Seager} \& {Mall{\'e}n-Ornelas}(2003)}]{Seager2003}
{Seager}, S., \& {Mall{\'e}n-Ornelas}, G. 2003, \apj, 585, 1038,
  \dodoi{10.1086/346105}

\bibitem[{{Seager} \& {Sasselov}(2000)}]{Seager2000}
{Seager}, S., \& {Sasselov}, D.~D. 2000, \apj, 537, 916, \dodoi{10.1086/309088}

\bibitem[{{Sing}(2010)}]{Sing2010}
{Sing}, D.~K. 2010, \aap, 510, A21, \dodoi{10.1051/0004-6361/200913675}

\bibitem[{{Sing} {et~al.}(2026){Sing}, {Lothringer}, {Valenti}, {Allen},
  {Bennett}, {Gasc{\'o}n}, {Mak}, {McCreery}, {Mukherjee}, {Ramos Rosado},
  {Schmidt}, {Stevenson}, {Thorngren}, \& {Wang}}]{Sing2026}
{Sing}, D.~K., {Lothringer}, J.~D., {Valenti}, J.~A., {et~al.} 2026, arXiv
  e-prints, arXiv:2609.00263, \dodoi{10.48550/arXiv.2609.00263}

\bibitem[{{Southworth}(2008)}]{Southworth2008}
{Southworth}, J. 2008, \mnras, 386, 1644,
  \dodoi{10.1111/j.1365-2966.2008.13145.x}

\bibitem[{{The Pandas Development Team}(2025)}]{pandas}
{The Pandas Development Team}. 2025, {pandas-dev/pandas: Pandas}, v2.3.1,
  Zenodo, \dodoi{10.5281/zenodo.3509134}

\bibitem[{{Triaud}(2018)}]{Triaud2018}
{Triaud}, A. H.~M.~J. 2018, in Handbook of Exoplanets, ed. H.~J. {Deeg} \&
  J.~A. {Belmonte}, 2, \dodoi{10.1007/978-3-319-55333-7_2}

\bibitem[{{Tsai} {et~al.}(2021){Tsai}, {Innes}, {Lichtenberg}, {Taylor},
  {Malik}, {Chubb}, \& {Pierrehumbert}}]{Tsai2021}
{Tsai}, S.-M., {Innes}, H., {Lichtenberg}, T., {et~al.} 2021, \apjl, 922, L27,
  \dodoi{10.3847/2041-8213/ac399a}

\bibitem[{{Tsai} {et~al.}(2023){Tsai}, {Lee}, {Powell}, {Gao}, {Zhang},
  {Moses}, {H{\'e}brard}, {Venot}, {Parmentier}, {Jordan}, {Hu}, {Alam},
  {Alderson}, {Batalha}, {Bean}, {Benneke}, {Bierson}, {Brady}, {Carone},
  {Carter}, {Chubb}, {Inglis}, {Leconte}, {Line}, {L{\'o}pez-Morales},
  {Miguel}, {Molaverdikhani}, {Rustamkulov}, {Sing}, {Stevenson}, {Wakeford},
  {Yang}, {Aggarwal}, {Baeyens}, {Barat}, {de Val-Borro}, {Daylan}, {Fortney},
  {France}, {Goyal}, {Grant}, {Kirk}, {Kreidberg}, {Louca}, {Moran},
  {Mukherjee}, {Nasedkin}, {Ohno}, {Rackham}, {Redfield}, {Taylor}, {Tremblin},
  {Visscher}, {Wallack}, {Welbanks}, {Youngblood}, {Ahrer}, {Batalha}, {Behr},
  {Berta-Thompson}, {Blecic}, {Casewell}, {Crossfield}, {Crouzet}, {Cubillos},
  {Decin}, {D{\'e}sert}, {Feinstein}, {Gibson}, {Harrington}, {Heng},
  {Henning}, {Kempton}, {Krick}, {Lagage}, {Lendl}, {Lothringer}, {Mansfield},
  {Mayne}, {Mikal-Evans}, {Palle}, {Schlawin}, {Shorttle}, {Wheatley}, \&
  {Yurchenko}}]{Tsai2023}
{Tsai}, S.-M., {Lee}, E. K.~H., {Powell}, D., {et~al.} 2023, \nat, 617, 483,
  \dodoi{10.1038/s41586-023-05902-2}

\bibitem[{Virtanen {et~al.}(2020)Virtanen, Gommers, Oliphant, Haberland, Reddy,
  Cournapeau, Burovski, Peterson, Weckesser, Bright, {van der Walt}, Brett,
  Wilson, Millman, Mayorov, Nelson, Jones, Kern, Larson, Carey, Polat, Feng,
  Moore, {VanderPlas}, Laxalde, Perktold, Cimrman, Henriksen, Quintero, Harris,
  Archibald, Ribeiro, Pedregosa, {van Mulbregt}, \& {SciPy 1.0
  Contributors}}]{scipy}
Virtanen, P., Gommers, R., Oliphant, T.~E., {et~al.} 2020, Nature Methods, 17,
  261, \dodoi{10.1038/s41592-019-0686-2}

\bibitem[{Ward(1963)}]{Ward1963}
Ward, J.~H. 1963, Journal of the American Statistical Association, 58, 236

\bibitem[{{Welbanks} {et~al.}(2024){Welbanks}, {Bell}, {Beatty}, {Line},
  {Ohno}, {Fortney}, {Schlawin}, {Greene}, {Rauscher}, {McGill}, {Murphy},
  {Parmentier}, {Tang}, {Edelman}, {Mukherjee}, {Wiser}, {Lagage}, {Dyrek}, \&
  {Arnold}}]{Welbanks2024}
{Welbanks}, L., {Bell}, T.~J., {Beatty}, T.~G., {et~al.} 2024, \nat, 630, 836,
  \dodoi{10.1038/s41586-024-07514-w}

\bibitem[{{Xue} {et~al.}(2024){Xue}, {Bean}, {Zhang}, {Welbanks}, {Lunine}, \&
  {August}}]{Xue2024}
{Xue}, Q., {Bean}, J.~L., {Zhang}, M., {et~al.} 2024, \apjl, 963, L5,
  \dodoi{10.3847/2041-8213/ad2682}

\end{thebibliography}
\bibliographystyle{aasjournal}

\appendix

\restartappendixnumbering
\renewcommand{\theHfigure}{appendix.\thefigure}
\renewcommand{\theHtable}{appendix.\thetable}

\section{Derivation of the theoretical amplification factor limit}\label{sec:appendix_sqrt}
Consider a transit light curve composed of three equal\vtwo{-}duration segments: a pre-transit baseline, an in-transit segment, and a post-transit baseline, each containing $N$ points. Let $\sigma_{\rm OOT}$ denote the typical per-point photometric uncertainty, assumed identical across all segments and uncorrelated.
The transit depth $D$ is measured as the difference between the out-of-transit mean flux and the in-transit mean flux. These two quantities are estimated independently from disjoint sets of points, so their uncertainties combine in quadrature. The uncertainty on the in-transit mean is $\sigma_{\rm OOT}/\sqrt{N}$, and the uncertainty on the out-of-transit mean, estimated from $2N$ baseline points, is $\sigma_{\rm OOT}/\sqrt{2N}$. The uncertainty on their difference is therefore
\begin{equation}
    \sigma_D = \sigma_{\rm OOT}\sqrt{\frac{1}{N} + \frac{1}{2N}} = \sigma_{\rm OOT}\sqrt{\frac{3}{2N}}.
\end{equation}
\vtwo{Thus}, with $\sigma_{\rm ref} \equiv \sigma_{\rm OOT}/\sqrt{N}$, the theoretical floor is $A_{\rm theory} = \sigma_D/\sigma_{\rm ref} = \sqrt{3/2}$. 

\vtwo{Notably, this value depends on the ratio of the out-of-transit baseline to the transit duration. Having 2:1 baselines (equal pre-, post-, and in-transit durations) is generally recommended (e.g., JWST/NIRCam\footnote{\url{https://jwst-docs.stsci.edu/jwst-near-infrared-camera/nircam-observing-strategies/nircam-time-series-observation-recommended-strategies}} observing guidelines) and gives the $\sqrt{3/2}$ limit derived above. In practice, though, a large fraction of single-transit observations have a 1:1 baseline (lacking pre- or post-transit data), whether by choice (e.g., JWST/NIRISS\footnote{\url{https://jwst-docs.stsci.edu/jwst-near-infrared-imager-and-slitless-spectrograph/niriss-example-science-programs/niriss-soss-time-series-observations-of-a-transiting-exoplanet/step-by-step-etc-guide-for-niriss-soss-time-series-observations-of-a-transiting-exoplanet}} observing guidelines) or due to observing constraints. In this case, repeating the above derivation returns $A_{\rm theory} = \sqrt{2}$. These two values bracket the range of theoretical floors one might reasonably quote depending on the observational baseline.}

\vtwo{We stress that none of the quantitative results presented in this paper are computed relative to $A_{\rm theory}$: it is specified purely so that readers have a reference for the amplification factor's noise-limited floor. Because our conclusions are insensitive to which of these values is adopted, we default to the 2:1 baseline value, $A_{\rm theory} = \sqrt{3/2}$, throughout the manuscript.}

\clearpage

\section{System and injection--retrieval parameters}\label{sec:appendix_table}

\autoref{tab:tab1} provides the parameters of the injection--retrieval framework used throughout the text.

\begin{deluxetable*}{lccccccc}[htbp]
\tabletypesize{\footnotesize}
\tablecaption{Injection--retrieval summary}
\tablehead{
\colhead{Parameter} & \multicolumn{7}{c}{Value}
}
\startdata
\multicolumn{8}{c}{\textit{Injection - System parameters}} \\
\hline
$R_p/R_\star$ & \multicolumn{7}{c}{$0.1$} \\
$i$ (degrees) & \multicolumn{7}{c}{$90$} \\
$P$ (days) & \multicolumn{7}{c}{$1$} \\
$a/R_\star$ & \multicolumn{7}{c}{$\sim4.21$ (Kepler's third law)} \\
$e$ & \multicolumn{7}{c}{$0$} \\
$\omega$ (degrees) & \multicolumn{7}{c}{$0$} \\
$T_0$ (days) & \multicolumn{7}{c}{$0$} \\
$R_\star$ ($R_{\odot}$) & \multicolumn{7}{c}{$1$} \\
$M_\star$ ($M_{\odot}$) & \multicolumn{7}{c}{$1$} \\
\hline
\multicolumn{8}{c}{\textit{Injection - LDCs}} \\
\hline
 & \autoref{sec:TDPP} & \autoref{sec:amp_factor_cause} & \multicolumn{5}{c}{\autoref{sec:boosting_science}} \\
 & illustrative & Global mode & Cluster 0 & Cluster 2 & Cluster 3 & Cluster 4 & Cluster 7\\
\hline
$c_1$ & $0.1$ & $0.6245$ & $0.5597$ & $0.7617$ & $0.6666$ & $0.5172$ & $0.6383$ \\
$c_2$ & $0.2$ & $-0.1898$ & $-0.1310$ & $-0.8502$ & $-0.8767$ & $-0.1913$ & $-0.0511$ \\
$c_3$ & $0.4$ & $0.1473$ & $0.4556$ & $1.4229$ & $0.8343$ & $0.0819$ & $-0.2722$ \\
$c_4$ & $0.3$ & $-0.0634$ & $-0.2398$ & $-0.4899$ & $-0.2968$  & $-0.0316$ & $0.1455$ \\ 
\hline
\multicolumn{8}{c}{\textit{Corresponding stellar parameters}} \\
\hline
& ---  & Global mode & Cluster 0 & Cluster 2 & Cluster 3 & Cluster 4 & Cluster 7 \\
\hline
$T_{\rm eff}$ (K) & --- & $8389$ & $4111$ & $4111$ & $4111$ & $5333$ & $5944$ \\
$\log g$ & ---  & $4.33$ & $3.67$  & $4.78$ & $3.89$ & $3.00$ & $4.33$ \\
$[{\rm M/H}]$ & ---  & $0.06$ & $-0.67$ & $0.78$ & $0.06$ & $0.06$ & $0.06$ \\
$\lambda$ ($\mu$m) & ---  & $0.86$ & $1.30$ & $0.71$ & $5.27$ & $3.11$ & $2.02$ \\
\hline
\hline
\multicolumn{8}{c}{\textit{Retrieval}} \\
\hline
Parameter & \multicolumn{3}{c}{Prior} & \multicolumn{4}{c}{Walker init.} \\
\hline
$R_P/R_\star$ & \multicolumn{3}{c}{$\mathcal{U}(0, 1)$} & \multicolumn{4}{c}{$\mathcal{U}(0.07, 0.15)$} \\
$i$ (degrees) & \multicolumn{3}{c}{$\mathcal{U}(70, 110)$} & \multicolumn{4}{c}{$\mathcal{U}(88, 92)$} \\
$P$ (days) & \multicolumn{3}{c}{$\mathcal{N}(1, 5\times10^{-4})$} & \multicolumn{4}{c}{$\mathcal{U}(0.9995, 1.0005)$} \\
$a/R_\star$ & \multicolumn{3}{c}{$\mathcal{U}(0, 50)$} & \multicolumn{4}{c}{$\mathcal{U}(a/R_\star-2, a/R_\star+2)$} \\
$\sqrt{e}\cos{\omega}$ & \multicolumn{3}{c}{$\mathcal{U}(-1, 1)$} & \multicolumn{4}{c}{$\mathcal{U}(-0.2, 0.2)$} \\
$\sqrt{e}\sin{\omega}$ & \multicolumn{3}{c}{$\mathcal{U}(-1, 1)$} & \multicolumn{4}{c}{$\mathcal{U}(-0.2, 0.2)$} \\
$c_i$ (uniform) & \multicolumn{3}{c}{$\mathcal{U}(-100, 100)$} & \multicolumn{4}{c}{$\mathcal{U}(c_i^{\rm true}-0.5, c_i^{\rm true}+0.5)$} \\
$c_i$ (Gaussian) & \multicolumn{3}{c}{$\mathcal{N}(c_i^{\rm true},\, f \cdot c_i^{\rm true})$} & \multicolumn{4}{c}{$\mathcal{U}(c_i^{\rm true}-0.5, c_i^{\rm true}+0.5)$} \\
$T_0$ (days) & \multicolumn{3}{c}{Fixed} & \multicolumn{4}{c}{---} \\
\enddata
\tablecomments{System parameters describe the hypothetical transit system used throughout the paper. Three sets of injection LDCs are reported, each used in a different section: (1) the illustrative set in \autoref{sec:TDPP} was chosen to systematically vary the order of the injected LDL (the $n^{\rm th}$-order law uses the first $n$ values, e.g., the second-order polynomial uses $[0.1, 0.2]$); (2) the global mode in \autoref{sec:amp_factor_cause} is the LDC vector lying closest to the global centroid of all points, representing the most typical regime of the physical LDC space; (3) the \vtwo{five} cluster modes in \autoref{sec:boosting_science} sample distinct regions of LDC space to assess how our conclusions generalize across stellar conditions. The stellar parameters reported correspond to the ($T_{\rm eff}$, $\log g$, $[{\rm M/H}]$, $\lambda$) of the intensity profile from which each mode's LDC vector was derived. Eccentricity and argument of periastron are fixed to circular values during injection but included as free retrieval parameters to characterize their effect on the posterior. The Gaussian LDC priors used in \autoref{sec:boosting_science} are centered on the injected truth with fractional width $f \in \{20, 10, 5, 1\}\%$ as a proxy for prior strength.}
\label{tab:tab1}
\end{deluxetable*}

\clearpage
\onecolumngrid

\section{Limb-darkening distribution and clustering}\label{sec:appendix_LD_choices}

\begin{figure}[htbp]
    \script{Appendix1_plot.py}
    \centering
    \includegraphics[width=\textwidth]{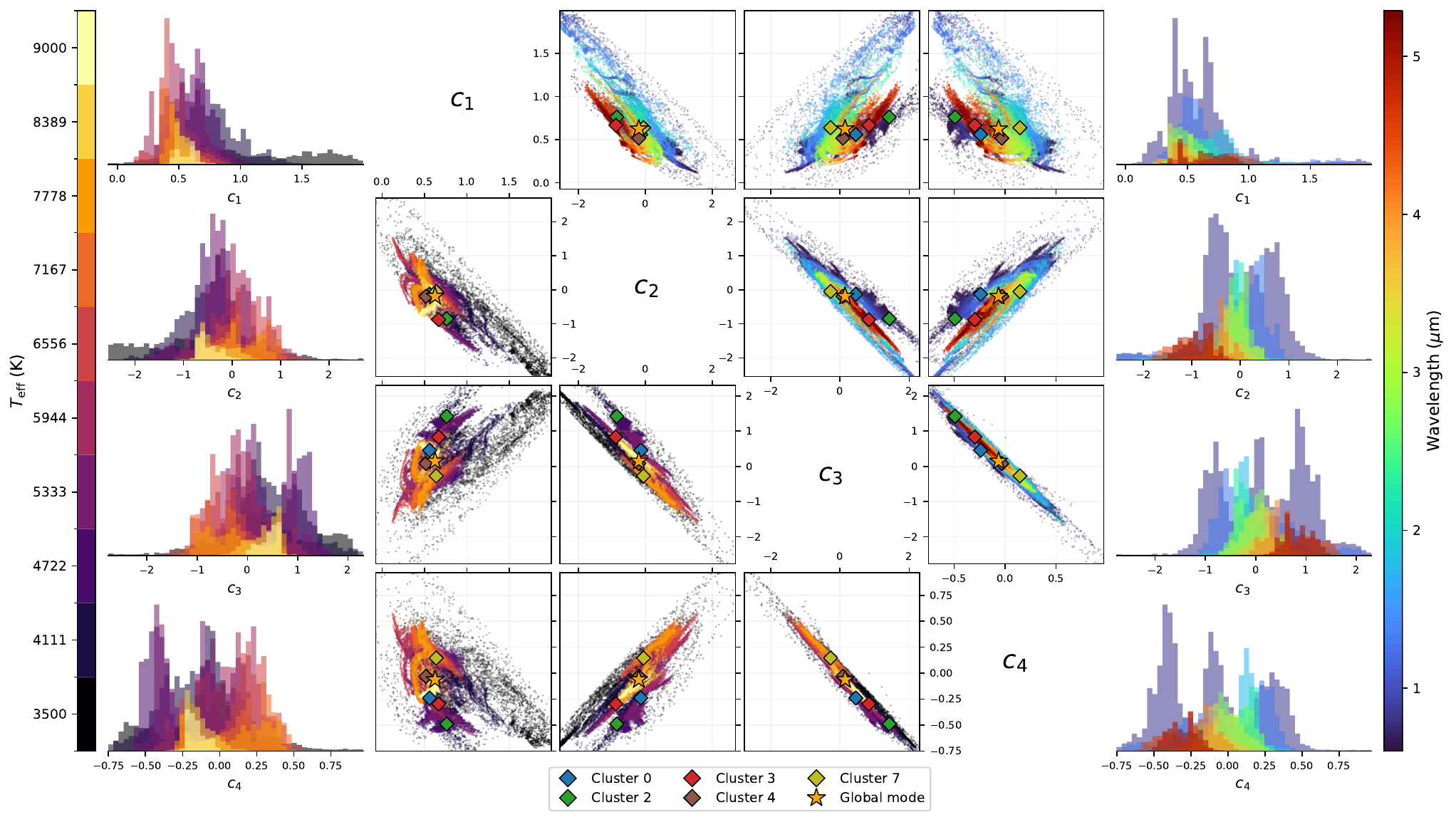}
    \caption{\textbf{Distribution of fourth-order nonlinear limb-darkening coefficients from the MPS-ATLAS set1 grid.} Double corner plot of the four LDCs ($c_1$, $c_2$, $c_3$, $c_4$) computed over $T_{\rm eff} \in [3500, 9000]$\,K, $\log g \in [3.0, 5.0]$, $[{\rm M/H}] \in [-5.0, 1.5]$, and JWST NIRSpec PRISM wavelengths. The \textit{lower-left} and \textit{upper-right} triangles show LDCs colored by $T_{\rm eff}$ and wavelength, respectively. The star and diamonds mark the location of the global and cluster modes used as injection LDCs throughout the manuscript. Diamond colors reflect the cluster colors used in \autoref{fig:appendix_fig3}. \textit{Left column:} 1D marginal histograms, grouped by the $T_{\rm eff}$ at which the LDCs were computed. \textit{Right column:} Same as the left column, but with grouping done according to wavelength. An animated version cycling through all four physical dimensions ($T_{\rm eff}$, $\log g$, $[{\rm M/H}]$, and wavelength) is available on GitHub \href{https://github.com/SamMerc/Transit-Information-Content/blob/main/src/static/corner_animation.gif}{\GitHubIcon}.}
    \label{fig:appendix_fig1}
\end{figure}

\begin{figure}[htbp]
    \script{Appendix2_plot.py}
    \centering
    \includegraphics[width=\textwidth]{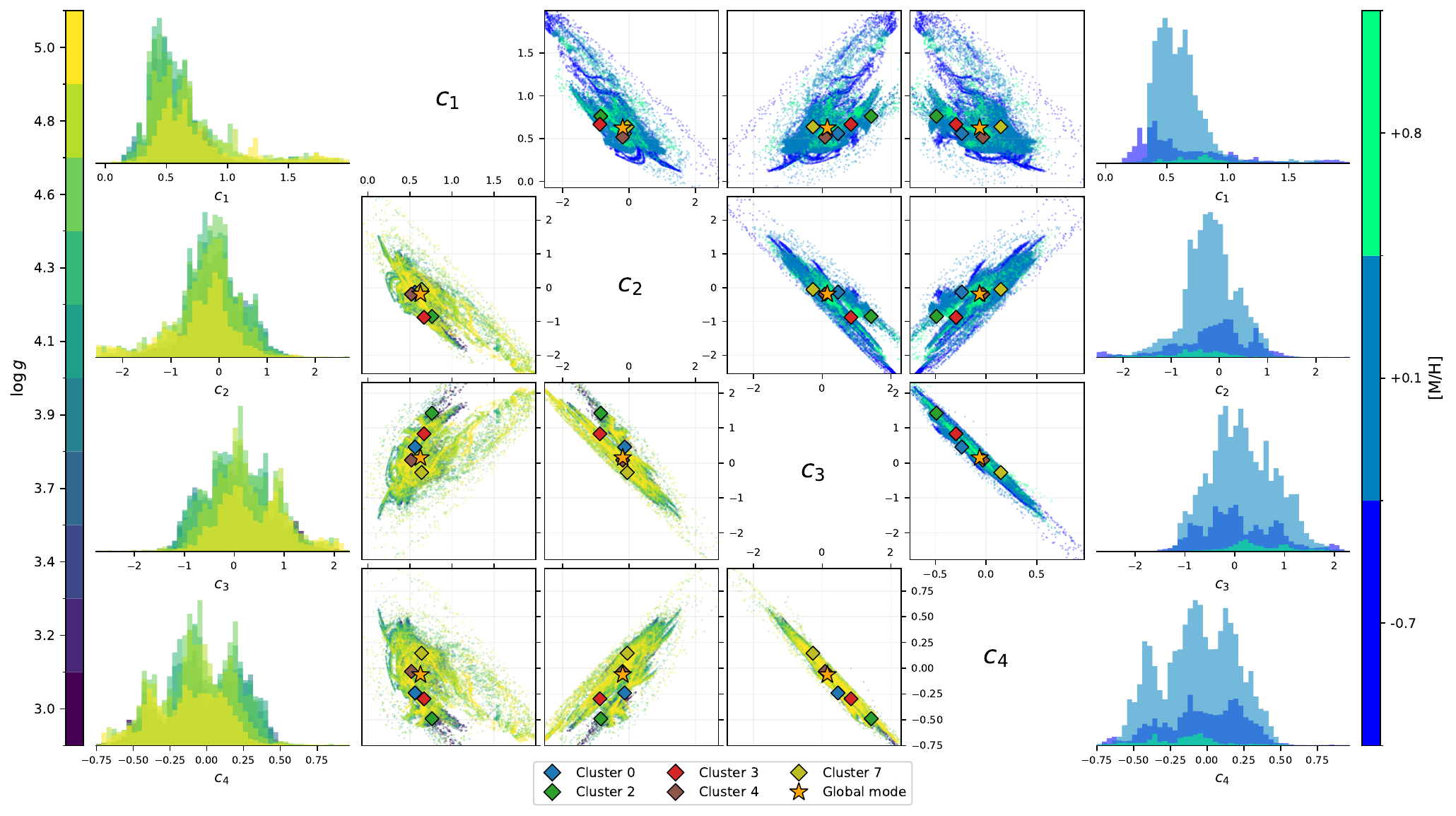}
    \caption{\textbf{Distribution of fourth-order nonlinear limb-darkening coefficients from the MPS-ATLAS set1 grid.} Same as \autoref{fig:appendix_fig1} except coloring LDCs by $\log g$ and $[{\rm M/H}]$.}
    \label{fig:appendix_fig2}
\end{figure}

\begin{figure}[htbp]
    \script{Appendix3_plot.py}
    \centering
    \includegraphics[width=\textwidth]{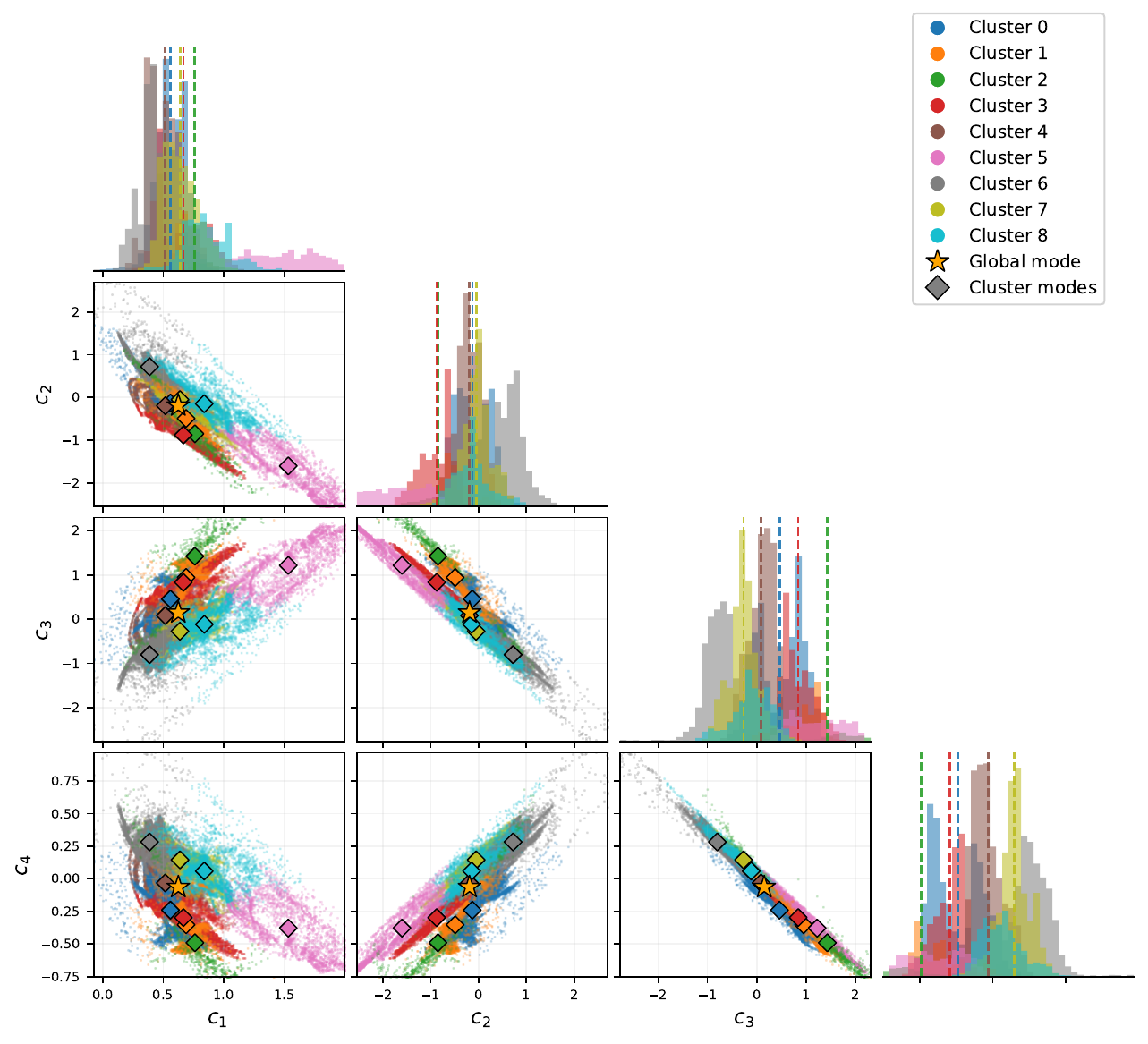}
    \caption{\textbf{Hierarchical clustering of the LDC space.} Corner plot of the four LDCs ($c_1$, $c_2$, $c_3$, $c_4$) colored by cluster membership as determined by the hierarchical clustering algorithm described in \autoref{sec:appropriate_coeff}. Each cluster is assigned a distinct color, with cluster modes shown as diamonds in the 2D panels. The global mode is depicted as an orange star. The diagonal panels show 1D marginal histograms of each coefficient, colored by cluster membership. Dashed vertical lines mark the modes for clusters 0, 2, 3, 4, and 7, which were used as injection--retrieval test cases in \autoref{fig:fig5} \vtwo{and \autoref{fig:appendix_fig6}}, and whose intensity profiles are shown in \autoref{fig:fig3}.}
    \label{fig:appendix_fig3}
\end{figure}

\autoref{fig:appendix_fig1} and \autoref{fig:appendix_fig2} \vtwo{show} the full distribution of fourth-order nonlinear LDCs across $T_{\rm eff}$, $\log g$, $[{\rm M/H}]$, and wavelength, while \autoref{fig:appendix_fig3} shows the same LDC space partitioned by hierarchical cluster membership.

\clearpage

\section{Sensitivity $\chi^2$ maps}\label{sec:appendix_sensitivity_heat_maps}

\begin{figure}[htbp]
    \script{Appendix4_plot.py}
    \centering
    \includegraphics[width=\textwidth]{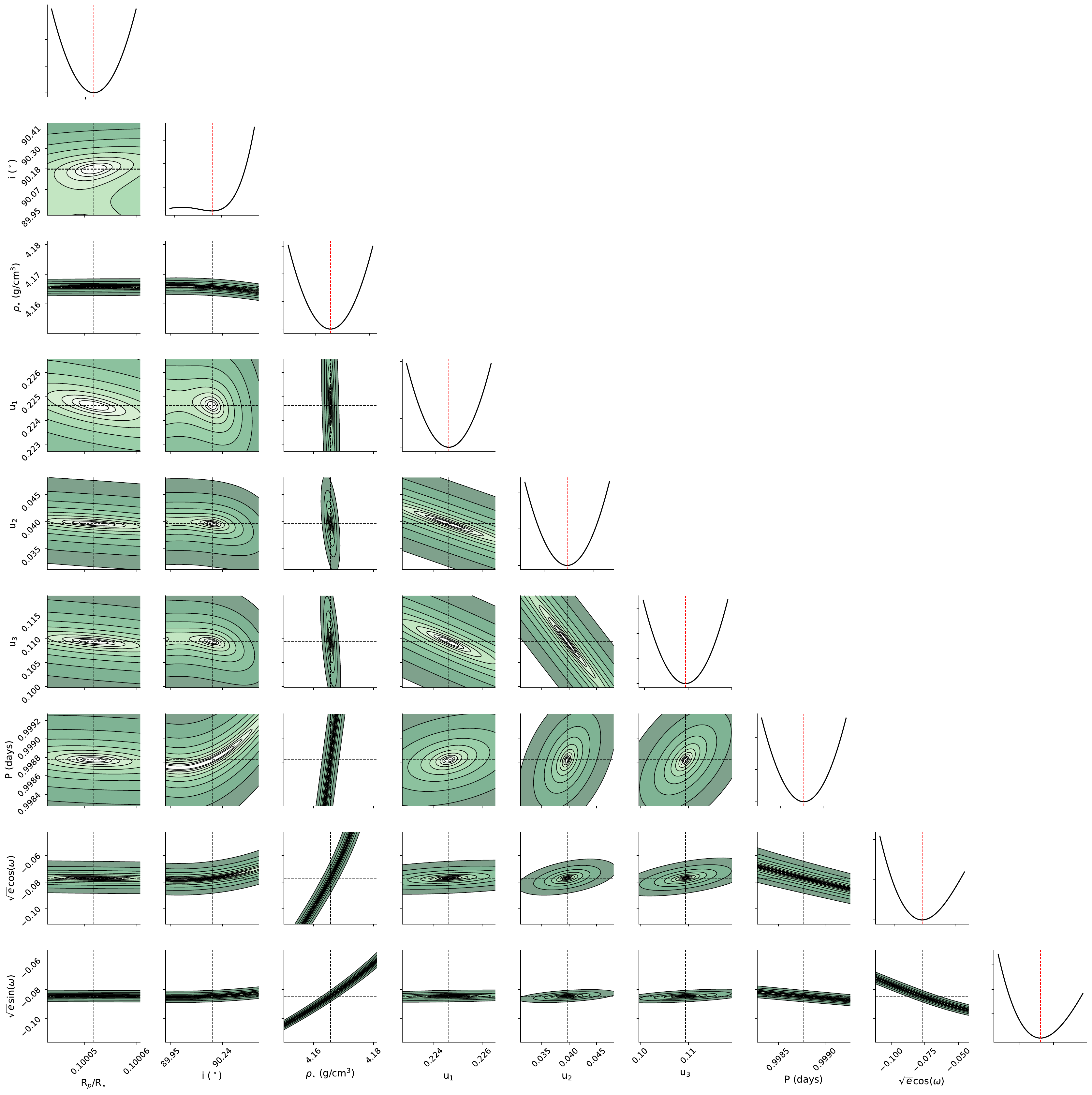}
    \caption{\textbf{$\chi^2$ maps across parameter pairs}. Corner plot of the $\chi^2$ maps used in our sensitivity test. Each diagonal subplot shows a 1-dimensional evaluation of the chi-squared space for the parameter corresponding to that column, where the vertical dashed red line indicates the truth value of that parameter. Off-diagonal subplots display two-dimensional iso-$\chi^2$ contours derived from the $\chi^2$ maps, where each map was obtained by varying the parameters indicated on the x and y axes, and the dashed black lines mark the truth values of the respective parameters. Finally, for certain contours that are not elliptical (e.g., $\sqrt{e}\sin{\omega}$ versus $P$)\vtwo{,} we approximate them as first-order polynomials and compute the rotation angle using $\theta = \mathrm{arctan(m)}$, where m is the slope.}
    \label{fig:appendix_fig4}
\end{figure}

\autoref{fig:appendix_fig4} presents the results of our sensitivity analysis.

\section{Comparison of MPS-ATLAS set1 and set2 LDCs}\label{sec:appendix_set1_set2}
\begin{figure}[htbp]
    \script{Appendix5_plot.py}
    \centering
    \includegraphics[width=0.93\textwidth]{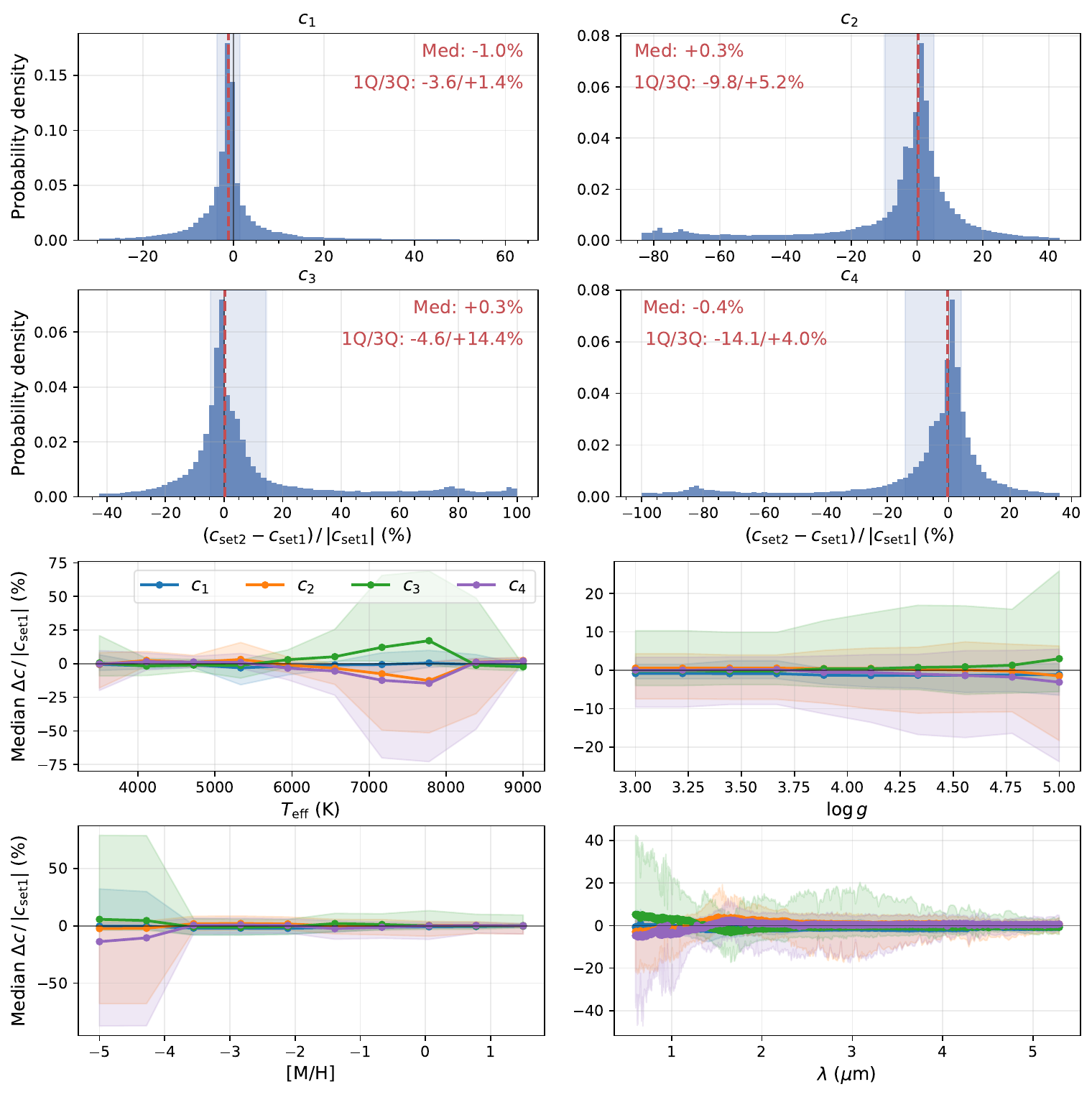}
    \caption{\textbf{Comparison of LDCs between the MPS-ATLAS set1 and set2 stellar atmosphere models.} \emph{Top four:} Histograms of the signed relative difference $(c_{\rm set2} - c_{\rm set1})\,/\,|c_{\rm set1}|$ for each of the four nonlinear LDL coefficients, computed point-by-point across matched grids of $T_{\rm eff}$, $\log g$, $[{\rm M/H}]$, and $\lambda$. Median values (red dashed lines) are near zero for all coefficients, confirming the negligible offset between the two model sets. The IQR grows with coefficient order, from $5\%$ for $c_1$ to $15\%$ for $c_2$ and $18$--$19\%$ for $c_3$ and $c_4$, reflecting the increasing difficulty in constraining higher-order coefficients that capture fine-scale intensity variations. \emph{Bottom four:} Median relative difference as a function of stellar parameters and wavelength, with shaded regions indicating the IQR. The inter-model spread exhibits a clear dependence on the physical parameters considered, underscoring that the IQRs derived from the top panels (and used to motivate the LDC priors in \autoref{sec:constraint_stellar_model}) are summary statistics that do not capture the full breadth of behavior across the parameter space.}
    \label{fig:appendix_fig5}
\end{figure}

\autoref{fig:appendix_fig5} summarizes the comparison of LDCs produced by the MPS-ATLAS set1 and set2 models and used in \autoref{sec:constraint_stellar_model} to establish the strength of priors that can be placed on LDCs from state-of-the-art stellar models.

\section{\vtwo{Injection--retrieval results for clusters 4 and 7}}\label{sec:appendix_fig6}

\begin{figure*}[htbp]
    \script{Appendix6_plot.py}
    \centering
    \includegraphics[width=0.85\textwidth]{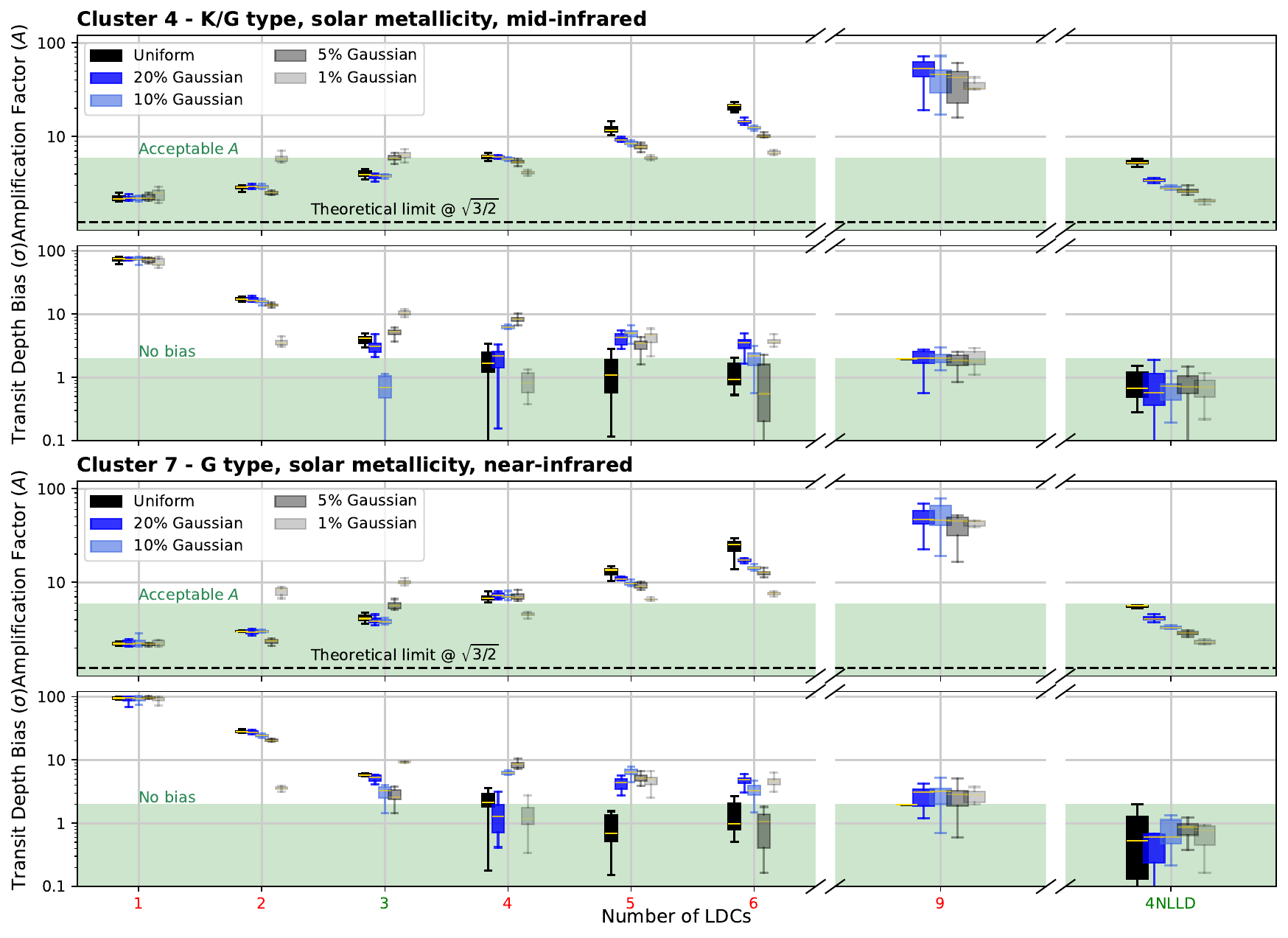}
    \caption{\textbf{Toward optimal limb-darkening \vtwo{parameterization} (continued).} \vtwo{Amplification factor (top panels) and transit depth bias (bottom panels) for the cluster 4 and cluster 7 injection modes, in the same format and with the same conventions as \autoref{fig:fig5}: results are shown as a function of the number of LDCs under limb-darkening priors ranging from uniform to Gaussian of varying width, with box plots summarizing 10 MCMC retrievals with different noise realizations. Green bands mark the regimes considered acceptable ($A \leq 6$; bias within $2\sigma$ of truth). These two clusters are discussed alongside the others in \autoref{sec:impact_prior_choice}}.}
    \label{fig:appendix_fig6}
\end{figure*}

\vtwo{For legibility, the injection--retrieval results of \autoref{sec:impact_prior_choice} are split across two figures. \autoref{fig:fig5} presents clusters 0, 2, and 3 in the main text and \autoref{fig:appendix_fig6} presents clusters 4 and 7 in the same format. The two are discussed together in \autoref{sec:impact_prior_choice}.}

\clearpage
\section{\vtwo{Chromatic dependence of the amplification factor}}\label{sec:appendix_A_lambda}

\begin{figure*}[htbp]
    \script{Appendix7_plot.py}
    \centering
    \includegraphics[width=\textwidth]{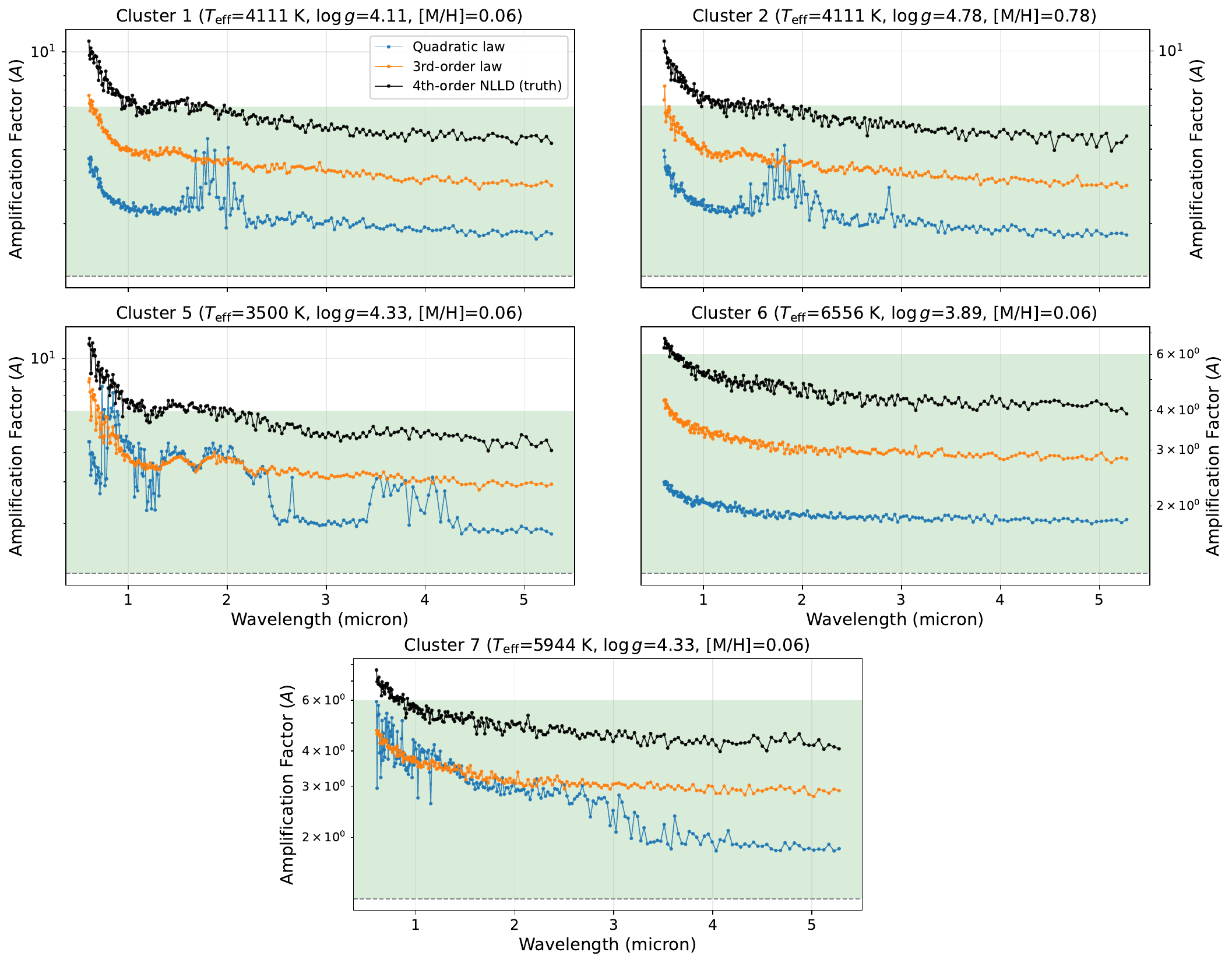}
    \caption{\vtwo{\textbf{Chromatic amplification factor incurred by the choice of LDL.} Amplification factor as a function of wavelength for the chromatic injection--retrievals of \autoref{sec:chrom_transit_depth_bias}, in the same format and with the same conventions as \autoref{fig:fig6}: each panel corresponds to a set of stellar parameters, labeled above it, and each light curve is retrieved with the quadratic (blue), third-order polynomial (orange) and injection (black) law. As in \autoref{fig:fig5}, the green band marks $A \leq 6$ and the dashed line $A$'s theoretical limit.}}
    \label{fig:appendix_fig7}
\end{figure*}

\vtwo{\autoref{fig:appendix_fig7} reports the amplification factor for the chromatic injection--retrievals of \autoref{sec:chrom_transit_depth_bias}, whose transit-depth biases are shown in \autoref{fig:fig6}. The two are discussed together in \autoref{sec:chrom_transit_depth_bias}.}

\clearpage
\section{\vtwo{Exploring the dependence of results on the choice of $\mu_{\rm min}$}}\label{sec:appendix_mu_min}

\begin{figure*}[htbp]
    \script{Appendix8_plot.py}
    \centering
    \includegraphics[width=0.85\textwidth]{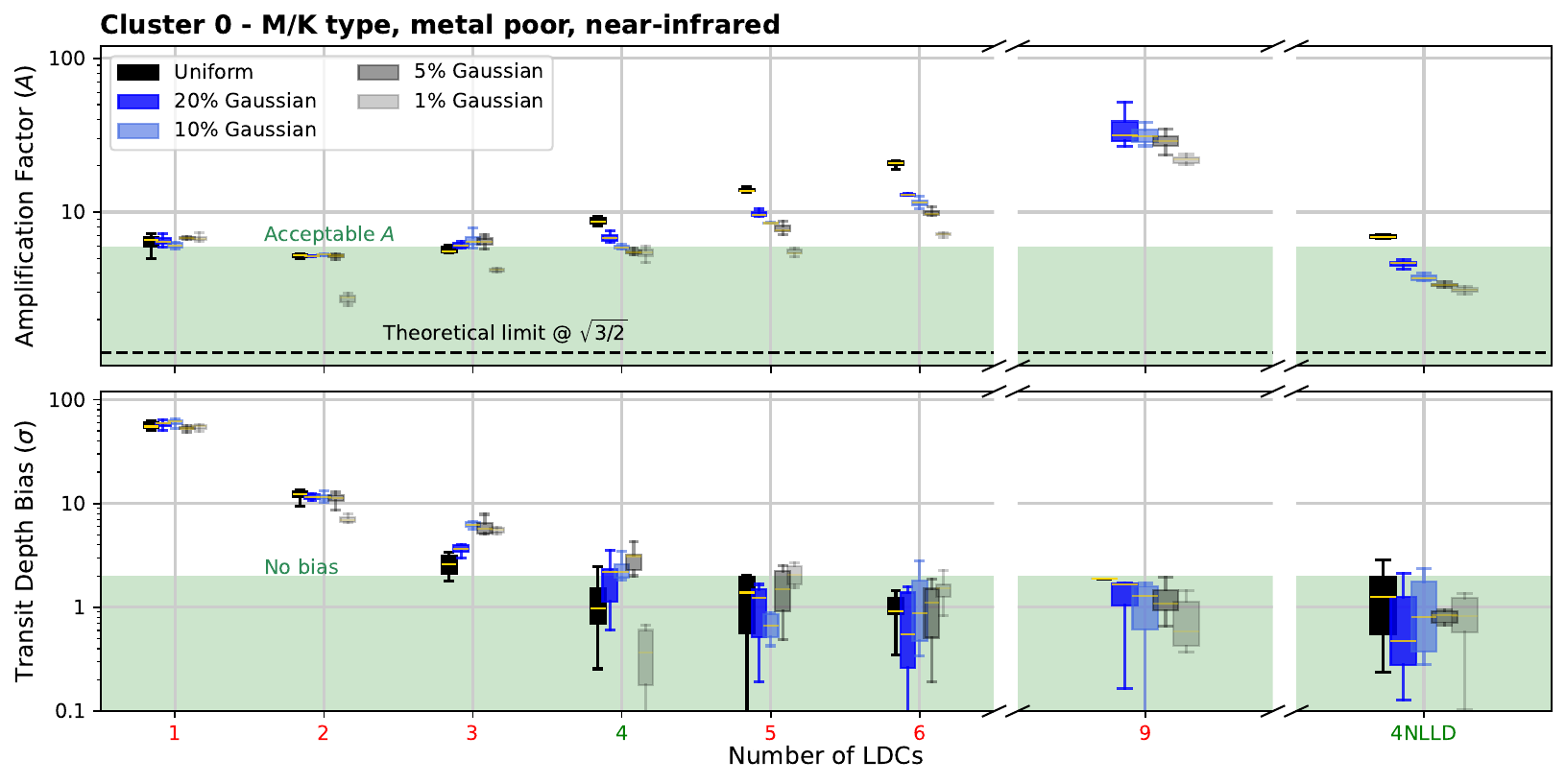}
    \caption{\vtwo{\textbf{Towards optimal limb-darkening parameterization, with a restricted $\mu$ grid.} Same format and conventions as \autoref{fig:fig5} for the cluster 0 mode, and with injection LDCs derived from a fit over a 100-point $\mu$ grid spanning $[0.2, 1.0]$ rather than the fiducial $[0.01, 1.0]$. Results are discussed in \autoref{sec:compare_empirics}.}}
    \label{fig:appendix_fig8}
\end{figure*}

\begin{figure*}[htbp]
    \script{Appendix9_plot.py}
    \centering
    \includegraphics[width=\textwidth]{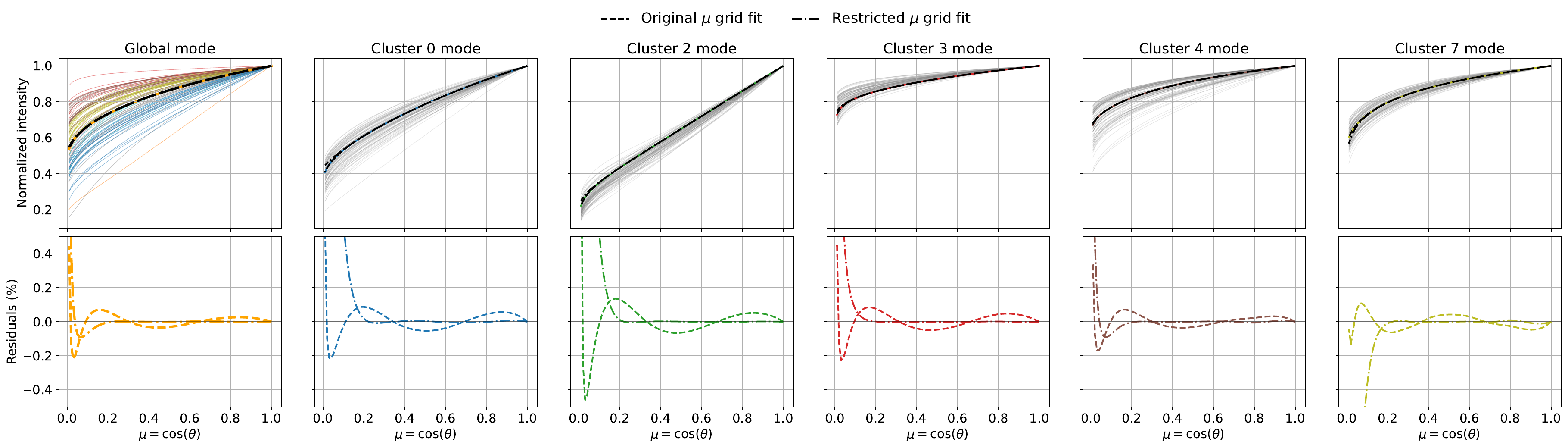}
    \caption{\vtwo{\textbf{Sensitivity of the extracted LDCs to the fitting $\mu$ range.} Shown are the same intensity profiles as \autoref{fig:fig3}, each fit with the fourth-order nonlinear LDL over two $\mu$ ranges: the fiducial grid spanning $[0.01, 1.0]$ (dashed lines) and the restricted grid spanning $[0.2, 1.0]$ (dashed-dotted lines). \emph{Top:} A random subset of 100 profiles is shown in the background of each panel, following the convention of \autoref{fig:fig3}. \emph{Bottom:} Relative difference residuals between each fit and the true intensity profile. Results are discussed in \autoref{sec:compare_empirics}.}}
    \label{fig:appendix_fig9}
\end{figure*}

\vtwo{\autoref{fig:appendix_fig9} compares the fiducial and restricted $\mu$-range fits to the representative intensity profiles, while \autoref{fig:appendix_fig8} shows the resulting impact on the injection--retrieval results for the cluster 0 mode.}

\end{document}